\documentclass[manuscript,revised]{geophysics}
\usepackage{latexsym,amsmath,amssymb,amsthm,systeme}
\usepackage[toc]{appendix}
\renewcommand{\figdir}{Fig}

\usepackage{setspace}
\usepackage{appendix}

\begin{document}

\title{Seismic Inversion by Multi-dimensional Newtonian Machine Learning}
\renewcommand{\thefootnote}{\fnsymbol{footnote}}

\ms{Seismic Inversion by Multi-dimensional Newtonian Machine Learning} 
\address{
\footnotemark[1] Deep Earth Imaging Future Science Platform, CSIRO, Kensington, Australia.
}
\author{Yuqing Chen\footnotemark[1] and Erdinc Saygin\footnotemark[1] }

\footer{Example}
\lefthead{Chen & Saygin}
\righthead{\emph{MNML inversion}}

\begin{abstract}

Newtonian machine learning (NML) is a wave-equation inversion method that inverts single-dimensional latent space (LS) features of the seismic data for retrieving the subsurface background velocity model. The single-dimensional LS features mainly contain the kinematic information of the seismic data, which are automatically extracted from the seismic signal by using an autoencoder network. Because its LS feature dimension is too small to preserve the dynamic information, such as the waveform variations, of the seismic data. Therefore the NML inversion is not able to recover the high-wavenumber velocity details. To mitigate this problem, we propose to invert multi-dimensional LS features, which can fully represent the entire characters of the seismic data. We denote this method as multi-dimensional Newtonian machine learning (MNML). In MNML, we define a new multi-variable connective function that works together with the multi-variable implicit function theorem to connect the velocity perturbations to the multi-dimensional LS feature perturbations. Numerical tests show that (1) the multi-dimensional LS features can preserve more data information than the single-dimensional LS features; (2) a higher resolution velocity model can be recovered by inverting the multi-dimensional LS features, and the inversion quality is comparable to that of FWI; (3) the MNML method requires a much smaller storage space than conventional FWI because only the low-dimensional representations of the high-dimensional seismic data are needed to be stored. The disadvantage of MNML is that it can more easily get stuck in local minima compared to the NML method. So we suggest a multiscale inversion approach that inverts for higher dimensional LS features as the iteration count increase. 

\end{abstract}

\section{Introduction}
Full waveform inversion (FWI) can recover a high-resolution subsurface model by minimizing the waveform difference between the observed and predicted seismic data \citep{tarantola1984inversion, virieux2009overview, warner2013anisotropic, perez2019velocity}. However, the FWI misfit function is highly nonlinear and characterized by many local minima. To mitigate these effects and acquire a reliable high-resolution image, a straightforward solution is to transform the complex data waveform to a simpler form. In this regard, \cite{luo1991wavea} inverted the background velocity model using wave equation traveltime inversion. To mitigate the cycle-skipping problem, \cite{bunks1995multiscale} proposed a multiscale inversion method that split the seismic data into different frequency ranges, inverting the low-frequency data first and then gradually inverting the higher-frequency information with an increase in the number of iterations. Instead of inverting the seismic waveforms, \cite{wu2014seismic} inverted the data envelope because it contains fewer complicated features than the original data waveform. They also reported that the seismic envelope contains stronger low-frequency information than the original waveform, which would increase the robustness of seismic inversion. \cite{warner2016adaptive} used a matching filter to transform the predicted data waveforms to the observed data waveforms. The velocity model can be updated by forcing the matching filter to be a zero-lag delta function. Moreover, a series of skeletonized inversion methods are developed and show many successful applications of inverting skeletonized data for various subsurface properties, such as velocity \citep{luo1991waveb} and quality factor Q \citep{dutta2016wave}. \cite{feng2019transmission} used the traveltimes of seismic reflection events to invert for the subsurface velocity and anisotropy parameters. \cite{li2016wave} recovered the S-wave velocity model by minimizing the dispersion residuals from the observed and predicted surface waves. \cite{liu20183d} and \cite{liu20203d} extended the dispersion inversion to 3D considering the cases for both flat and irregular surfaces. \cite{dutta2016wave} used the peak frequency shift information of early arrivals to extract the subsurface Qp model. Similarly, \cite{li2017wave} found the optimal Qs model using the peak-frequency shifts between the observed and predicted surface waves.

The above-mentioned inversion methods which extract the simplified data information, or skeletonized information, from seismic waveform are based on human knowledge and experience. Recently, machine learning provides opportunities for automatically extracting the skeletonized information from seismic data without human interference. Machine learning can identify important features and patterns from the data, and then a wave equation methodology can be used to invert these patterns for the subsurface velocity model. 

Pattern recognition has been widely used in many fields, such as computer vision, speech, and face recognition. In the seismic domain, \cite{valentine2012data} showed that an unsupervised neural network, such as an autoencoder, can be trained to learn a non-linear transformation that transforms the high-dimensional seismic traces to its low-dimensional representations. These low-dimensional representations preserve critical information contained in the seismic traces. \cite{chen2020seismic} showed that the critical information is related to the kinematic information of the seismic data, such as traveltimes, for one-dimensional LS features, and suggested that the one-dimensional LS features can be used to recover the subsurface background velocity model. However, the high-wavenumber velocity details are missing in their inversion because the one-dimensional LS features are not able to preserve the dynamic information of the seismic data, such as the waveform variations.

\begin{figure}[h]
\centering
\includegraphics[width=1\columnwidth]{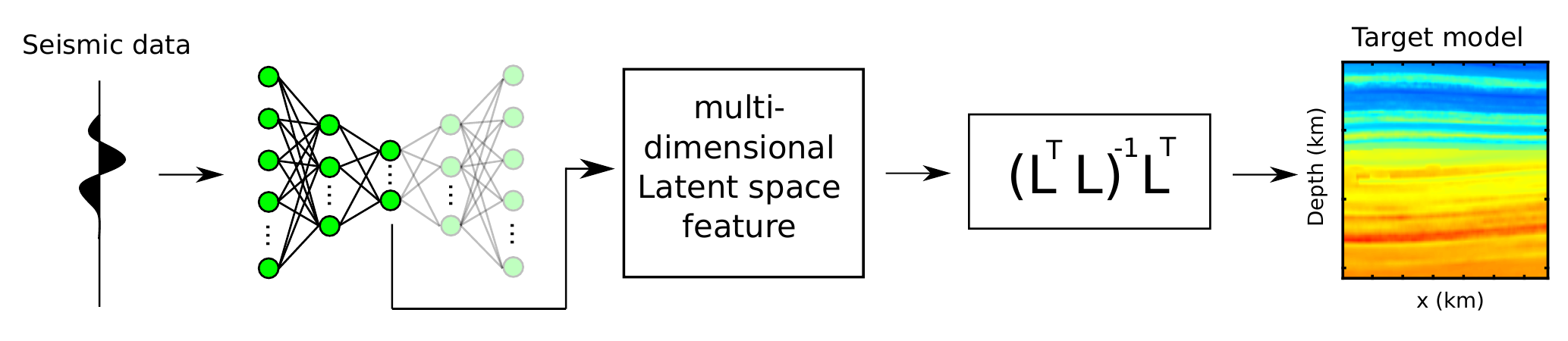}
\caption{The high-level strategy of the multi-dimensional Newtonian machine learning inversion. The multi-dimensional latent space features are automatically extracted from the seismic waveform by an autoencoder network, which are then inverted by wave equation inversion kernel for the subsurface velocity model.}
\label{fig:workflow1}
\end{figure}

To recover the high-wavenumber velocity details, we now present a multi-dimensional Newtonian machine learning (MNML) method that uses an autoencoder with a multi-dimensional LS to preserve the entire content of the seismic data. The high-level strategy of MNML inversion is shown in Figure \ref{fig:workflow1}, where $\mathbf{L}$ and $\mathbf{L}^{T}$ correspond to the forward and adjoint modeling operator. The MNML misfit function computes the spatial distance of the multi-dimensional LS features between the observed and predicted data in a multi-dimensional LS. This spatial distance quantifies the accuracy of the inverted model, which becomes zero when the inverted model is close enough to the true model. As there are no partial differential equations (PDE) that contains both the LS feature and the velocity parameter terms, so we define a multi-variable connective function that works together with the multi-variable implicit function theorem \citep{Guo2016multi} to compute the derivative between the LS features and the velocity parameters. The gradient formula of the MNML method is represented as a $n \times n$ matrix, where $n$ indicates the dimension of the LS vector. Numerical tests on both synthetic and real data demonstrate that the MNML method can effectively recover the high-wavenumber velocity model with a spatial resolution similar to a FWI tomogram. Moreover, MNML demands much less memory storage than FWI because the multi-dimensional LS features are at least 100 times smaller than the original seismic data. But a disadvantage of MNML inversion compared to the NML and conventional skeletonized inversion methods is that it is easier to getting stuck in local minima, because the multi-dimensional LS features are associated with many more events in the seismic traces than the single-dimensional LS features. To mitigate this problem, we propose a multiscale inversion approach which first uses NML inversion to recover the background velocity model, then uses MNML to reconstruct the high-wavenumber velocity details. 

In the following sections, we first describe the theory of autoencoders and then illustrate the effect of the LS dimension on the data reconstruction errors of autoencoder. Next, we introduce the theory of the MNML method and finally present both synthetic and field data tests. We conclude the MNML method and propose our future research in the last section.

\section{Theory}

\subsection{Autoencoder}
\begin{figure}[h]
\centering
\includegraphics[width=1\columnwidth]{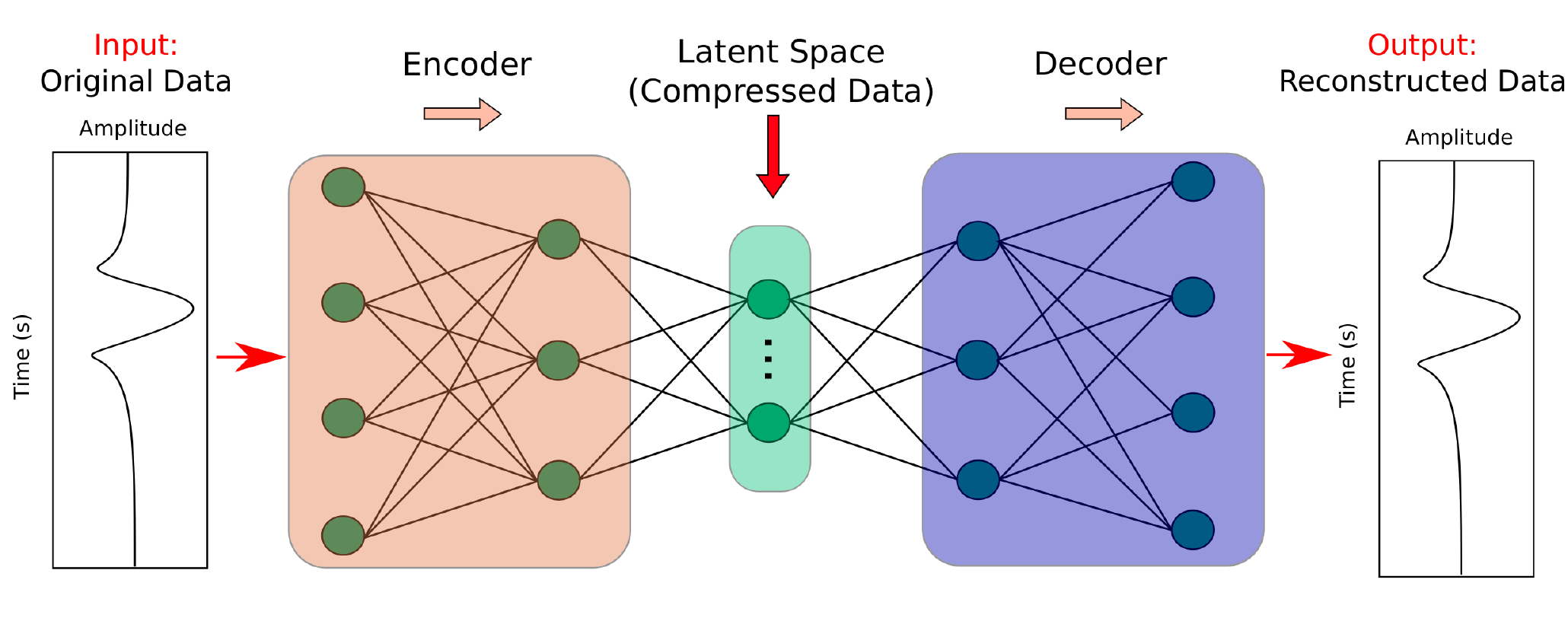}
\caption{An example of a three-layer architechture with a multi-dimentional latent space.}
\label{fig:AE}
\end{figure}

An example of a three-layer autoencoder architecture is shown in Figure \ref{fig:AE}, which includes an encoder network, latent space, and decoder network. The pink box emphasizes the encoder network that compresses the input data to a lower-dimensional space by using several neural layers. The number of neurons in each layer gradually decreases. The green box represents the latent space which contains the lowest-dimensional representation of the input data. The compressed data is then passed to the decoder network (purple box) to get the decoded waveform that has the same dimension as the input data. The decoder network is often the mirror of the encoder network which is composed of several neural layers with an increasing number of neurons in each layer. The misfit function of the autoencoder $\epsilon=(d_{input}-d_{decode})^{2}$ computes the differences between the input $d_{input}$ and decoded data $d_{decode}$, and the training process stops when the misfit reduces to a certain threshold. For a well-trained autoencoder, it should be able to extract the effective low-dimensional representation of the high-dimensional seismic data.

\subsection{Reconstruction ability of autoencoder}
\begin{figure}
\centering
\includegraphics[width=0.8\columnwidth]{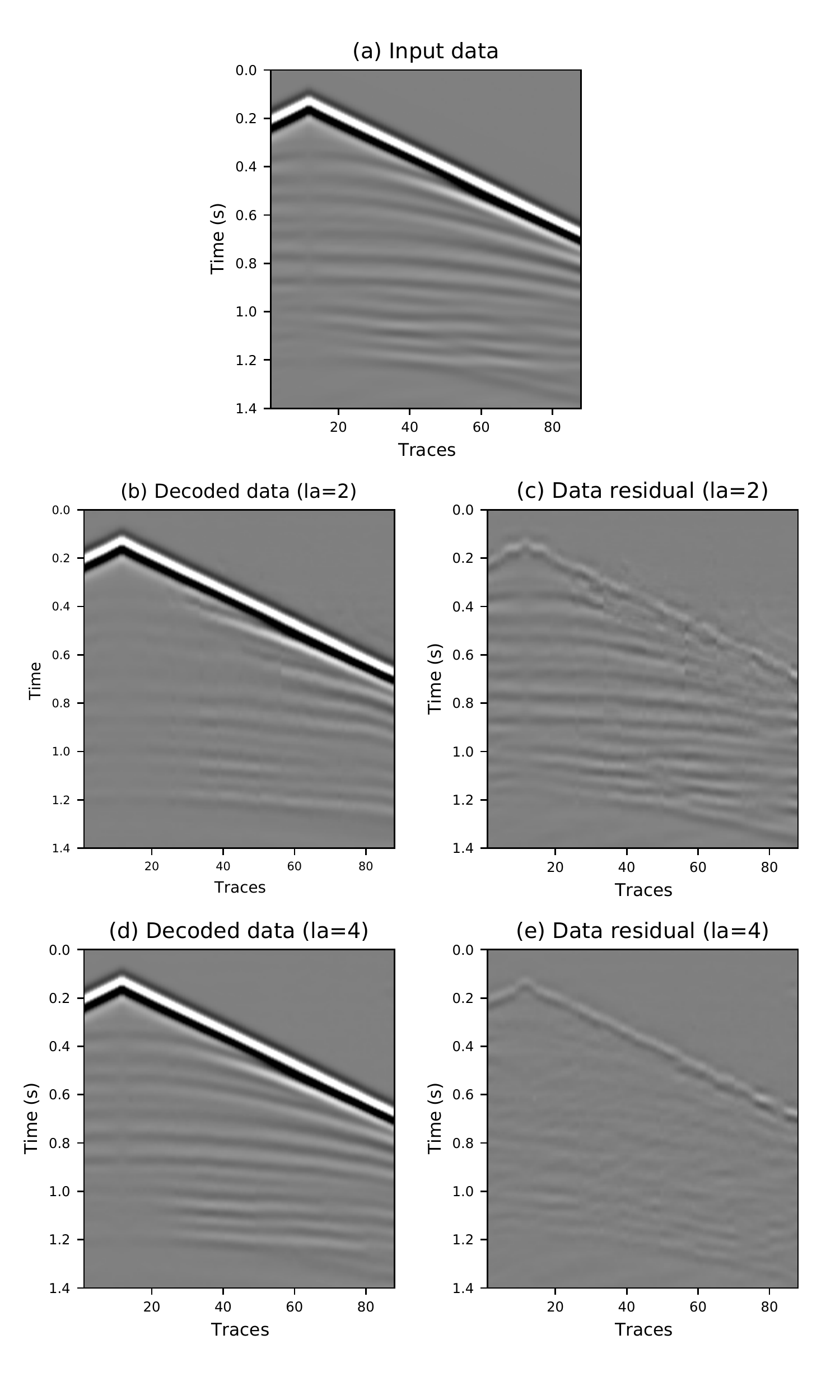}
\caption{(a) The input data, which is a subset of the training data. (b) The decoded waveform of the autoencoder with a two-dimensional latent space and (c) its differences with the input data. (d) The decoded waveform of the autoencoder with a four-dimensional latent space and (e) its difference with the input data.}
\label{fig:recon1}
\end{figure}

The reconstruction ability of an autoencoder is affected by several factors, such as the number of neural layers, the number of neurons in each layer, and the LS dimension. Here we only discuss and analyze how the LS dimension affects the reconstruction ability of an autoencoder. We design two autoencoders with the same encoder and decoder network, where each consists of multi-perceptrons. We set one of the autoencoder's LS dimensions to two (la=2) and another one to four (la=4). Two thousand seismic traces are used to train these two autoencoders, and a subset of the training data is shown in Figure \ref{fig:recon1}a. Figure \ref{fig:recon1}b shows the decoded waveform from the autoencoder with a two-dimensional LS. It shows that the first arrival energy has been well recovered, but some reflection events are missing. Because the two-dimensional LS is too small to preserve the reflection energies. These missing reflection events can be clearly seen in Figure \ref{fig:recon1}c which shows the data difference between the input and decoded data. Figure \ref{fig:recon1}d shows the decoded waveform from the autoencoder with a four-dimensional LS, where both the first-arrival and reflection energies have been well recovered. Its data residual given in Figure \ref{fig:recon1}e shows that the differences of the reflection events are largely minimized compared to Figure \ref{fig:recon1}c. Therefore we conclude that an autoencoder with a larger LS dimension can better preserve and reconstruct input data information than an autoencoder with a smaller LS. 

\begin{figure}
\centering
\includegraphics[width=0.85\columnwidth]{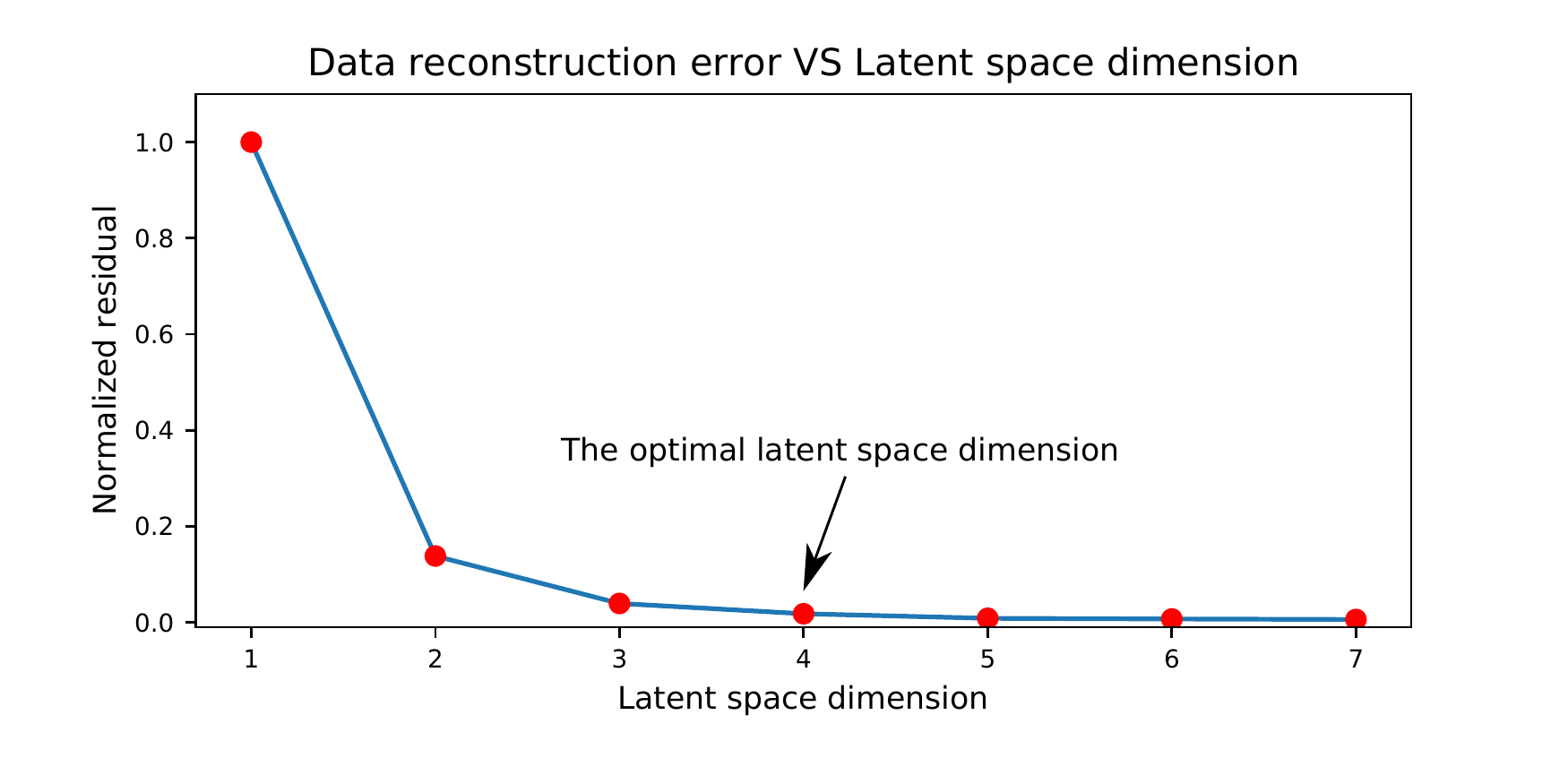}
\caption{The normalized data reconstruction error versus latent space dimension for the example given in Figure \ref{fig:recon1}. The black arrow points to the optimal latent space dimension. Identifying the optimal dimension as the one where the reconstruction error shows little decrease is known as the elbow method \citep{Schuster2017}.}
\label{fig:recon2}
\end{figure}

However, the data reconstruction ability of an autoencoder increases slowly when the LS dimension reaches a certain level. Figure \ref{fig:recon2} shows a curve that describes the data reconstruction error versus the LS dimension changes for this example. It shows that the reconstruction error drops rapidly when the LS dimension goes from one to four. But when the LS dimension is larger than four, the reconstruction error barely changes. This phenomenon suggests that the four-dimensional LS is good enough to preserve and reconstruct the entire information of the input data, and this four-dimensional LS is considered as the optimal LS dimension.  

\subsection{Misfit Function}
\begin{figure}
\centering
\includegraphics[width=0.8\columnwidth]{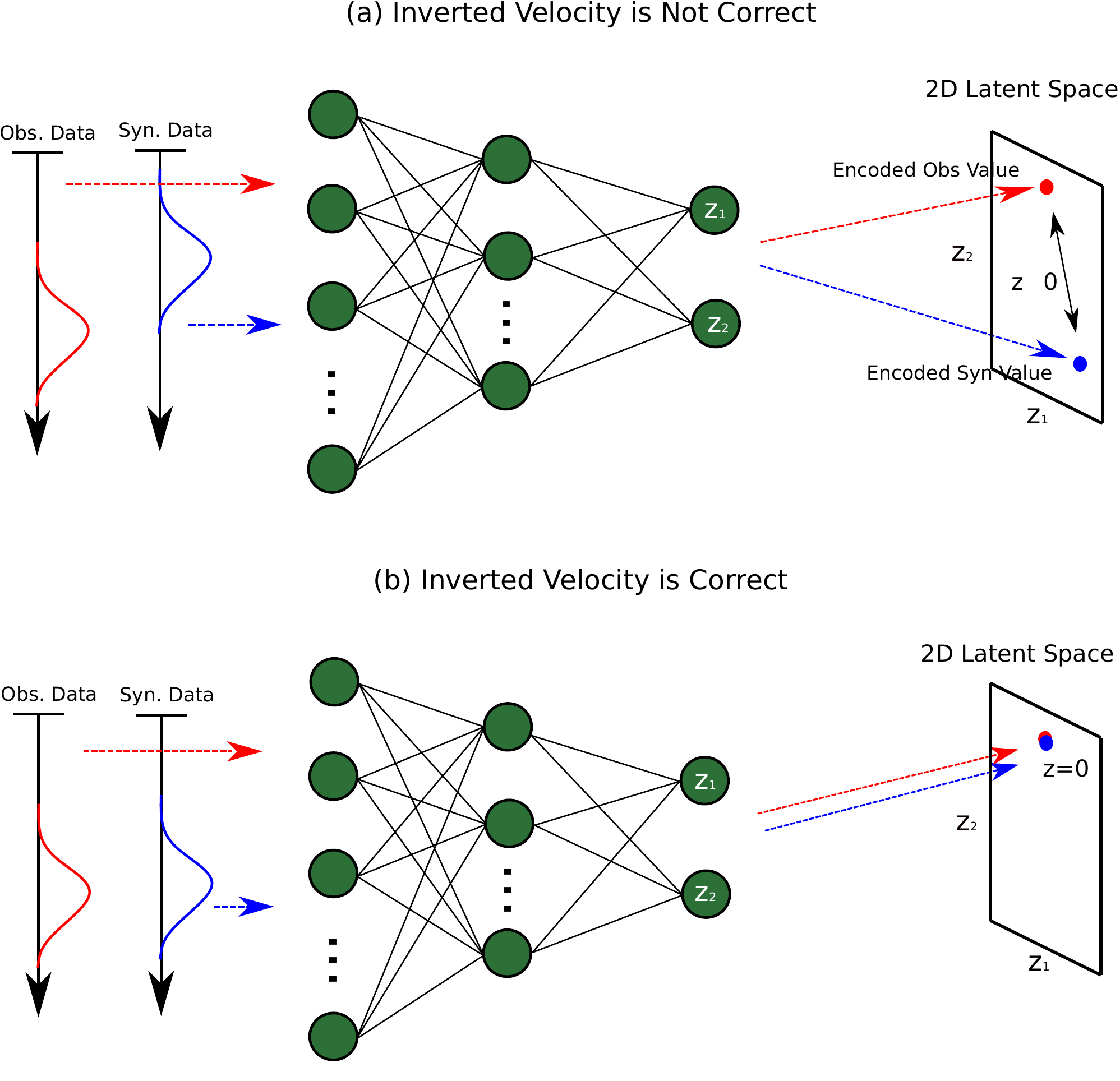}
\caption{(a) If the inverted velocity is incorrect, the LS feature residual norm is large, otherwise (b) the norm of the LS residual is small when the predicted and observed LS points are close together.}
\label{fig:demon2}
\end{figure}

A $nt \times 1$ seismic trace can be viewed as a data point in a $nt$ dimensional space, where $nt$ is the number of time samples of a seismic trace. Therefore the FWI misfit function measures the spatial distance between the observed and predicted data point in an $nt$ dimensional space. Similarly, the MNML method measures the spatial distance of the LS feature of the observed and predicted trace in a $n \times 1$ dimensional space, where $n$ indicates the dimensionality of the LS features which is much smaller than $nt$. Figure \ref{fig:demon2} illustrates a keystone idea underlying the MNML method. An autoencoder compresses the high-dimensional observed and predicted seismic trace to a two-dimensional LS. The spatial distance between the compressed observed and predict two-dimensional LS features can be used to quantify the accuracy of the inverted model. A large distance means that the inverted model is far away from the true model, but a small distance indicates a strong similarity between the true and inverted model. In general, for the MNML method, the LS dimension could be an arbitrary number that is larger than one and smaller than $nt$. However, we can find the optimal LS dimension $n$ according to the relationship between the LS dimension $n$ and data reconstruction error of the autoencoder, which is demonstrated in Figure \ref{fig:recon2}. We identify the optimal dimension as the 'elbow' point of the curve which shows a relatively small decrease in data reconstruction error	with increasing dimension.

The misfit function of the MNML method can be written as 
\begin{equation}
\epsilon = \sum_{s}\sum_{r}\sum_{k} \Delta z_{k}(\mathbf{x}_{r},\mathbf{x}_{s})^{2},
\label{eq:misift1}
\end{equation}
where $\Delta z$ represents the multi-dimensional LS feature difference of the observed and predicted data. The locations of the source and receiver are represented by $\mathbf{x}_{s}$ and $\mathbf{x}_{r}$, respectively, where $s$ and $r$ indicate the source and receiver index. Here, $k$ indicates the dimension index of the multi-dimensional LS feature. For simplicity, we assume the LS feature dimension equals two. Then equation \ref{eq:misift1} can be re-written as 
\begin{equation}
\epsilon = \sum_{s} \sum_{r} \big[\Delta z_{1}(\mathbf{x}_{r},\mathbf{x}_{s})^{2}+ \Delta z_{2}(\mathbf{x}_{r},\mathbf{x}_{s})^{2} \big].
\label{eq:misfit2}
\end{equation}
The gradient $\gamma (\mathbf{x})$ can be found by taking the derivative of the misfit $\epsilon$ to the velocity $v(\mathbf{x})$ as
\begin{equation}
\gamma(\mathbf{x})=-\frac{\partial\epsilon}{\partial v(\mathbf{x})}=-\sum_{s}\sum_{r}\bigg[ \bigg( \frac{\partial \Delta z_{1}(\mathbf{x}_{r},\mathbf{x}_{s})}{\partial v(\mathbf{x})}\bigg)^{T}\Delta z_{1}(\mathbf{x}_{r},\mathbf{x}_{s})  + \bigg( \frac{\partial \Delta z_{2}(\mathbf{x}_{r},\mathbf{x}_{s})}{\partial v(\mathbf{x})} \bigg)^{T}\Delta z_{2}(\mathbf{x}_{r},\mathbf{x}_{s}) \bigg].
\label{eq:grad1}
\end{equation}
Equation \ref{eq:grad1} shows that the velocity gradient calculation is carried out at each LS dimension.

\subsection{Connective Function}
To compute the gradient $\gamma(\mathbf{x})$, we need to define the relationship between the velocity perturbation and the multi-dimensional LS feature perturbation mathematically. To do so, we build a multi-variable connective function that measures the similarity between the observed and predicted seismic data through cross-correlation:
\begin{equation}
F(\tilde{z}_{1},\tilde{z}_{2};v)=\int p_{(z_{1}-\tilde{z}_{1},z_{2}-\tilde{z}_{2})}^{obs}(\mathbf{x}_{r}, t; \mathbf{x}_{s})p_{(z_{1},z_{2})}^{pred}(\mathbf{x}_{r},t;\mathbf{x}_{s}) dt,
\label{eq:connect1}
\end{equation}
where $p_{(z_{1},z_{2})}^{pred}(\mathbf{x}_{r},t;\mathbf{x}_{s})$ indicates a predicted seismic trace recorded at the receiver location $x_{r}$ for a source at $\mathbf{x}_{s}$. The subscripts $z_{1}$ and $z_{2}$ represent the LS feature values of this trace for the first and second LS dimensions, respectively. Similarly, $p_{(z_{1}-\tilde{z}_{1},z_{2}-\tilde{z}_{2})}^{obs}(\mathbf{x}_{r}, t; \mathbf{x}_{s})$ represents the observed seismic trace at the same source-receiver location with the LS feature value equal to $(z_{1}-\tilde{z}_{1},z_{2}-\tilde{z}_{2})$, where $(\tilde{z}_{1},\tilde{z}_{2})$ is the difference between the observed and predicted multi-dimensional LS features. This difference will be zero ($(\tilde{z}_{1},\tilde{z}_{2})=(0,0)$) if the velocity model we build is accurate enough, otherwise $(\tilde{z}_{1},\tilde{z}_{2}) \neq (0,0)$. We need to (1) find the optimal LS feature difference $(\tilde{z}_{1},\tilde{z}_{2})=(\Delta z_{1},\Delta z_{2})$ that maximize equation \ref{eq:connect1}; then (2) recover the accurate velocity model by minimizing the difference. The optimal difference can be found by setting equation $\nabla F(\tilde{z}_{1},\tilde{z}_{2}) |_{(\tilde{z}_{1}=\Delta z_{1}, \tilde{z}_{2}=\Delta z_{2})}=0$ as
\begin{align}
\nabla F(\tilde{z}_{1},\tilde{z}_{2}) |_{(\tilde{z}_{1}=\Delta z_{1}, \tilde{z}_{2}=\Delta z_{2})} &=(\frac{\partial F}{\partial \tilde{z}_{1}},\frac{\partial F}{\partial \tilde{z}_{2}}) \nonumber \\
&=\int p_{(z_{1},z_{2})}^{syn}(\mathbf{x}_{r},t;\mathbf{x}_{s})\bigg (\frac{p_{(z_{1}-\Delta z_{1},z_{2}-\Delta z_{2})}^{obs}}{\partial \tilde{z}_{1}}, \frac{p_{(z_{1}-\Delta z_{1},z_{2}-\Delta z_{2})}^{obs}}{\partial \tilde{z}_{2}} \bigg) dt = 0, \label{eq:connect2}
\end{align}
Rewriting equation \ref{eq:connect2} in matrix form and combining it with the multi-variable implicit function theorem \citep{krantz2012implicit, Guo2016multi} , we can get the Fr\'echet derivatives
\begin{equation}
\begin{bmatrix}
\frac{\partial \Delta z_{1}}{\partial v} \\ \frac{\partial \Delta z_{2}}{\partial v}
\end{bmatrix}
=
\begin{bmatrix}
\frac{\partial^{2} F}{\partial \tilde{z}_{1}^{2}} & \frac{\partial^{2} F}{\partial \tilde{z}_{1} \partial \tilde{z}_{2}} \\
\frac{\partial^{2} F}{\partial \tilde{z}_{1} \partial \tilde{z}_{2}} & \frac{\partial^{2} F}{\partial \tilde{z}_{2}^{2}}
\end{bmatrix}^{-1}
\begin{bmatrix}
\frac{\partial^{2} F}{\partial \tilde{z}_{1} \partial v} \\ \frac{\partial^{2} F}{\partial \tilde{z}_{2} \partial v}
\end{bmatrix}.
\label{eq:connect3}
\end{equation}
The detailed derivations from equations \ref{eq:connect2} and \ref{eq:connect3}, and the formulas for the partial derivatives on the right-hand-side of equation \ref{eq:connect3} can be found in the Appendix B.

\subsection{Gradient}

Substituting equation \ref{eq:connect3} into equation \ref{eq:grad1}, we get the gradient formula for the two-dimensional NML 

\begin{align}
\gamma(\mathbf{x}) & = -\sum_{s}\sum_{r} \Bigg \langle
\begin{bmatrix}
\frac{\partial^{2} F}{\partial \tilde{z}_{1}^{2}} & \frac{\partial^{2} F}{\partial \tilde{z}_{1} \partial \tilde{z}_{2}} \\
\frac{\partial^{2} F}{\partial \tilde{z}_{1} \partial \tilde{z}_{2}} & \frac{\partial^{2} F}{\partial \tilde{z}_{2}^{2}}
\end{bmatrix}^{-1}
\begin{bmatrix}
\frac{\partial^{2} F}{\partial \tilde{z}_{1} \partial v} \\ \frac{\partial^{2} F}{\partial \tilde{z}_{2} \partial v}
\end{bmatrix},  
\begin{bmatrix}
\Delta z_{1} \\ \Delta z_{2}
\end{bmatrix} \Bigg \rangle.
\label{eq:grad}
\end{align}
If we assume $z_{1}$ and $z_{2}$ are weakley correlated, then $ \frac{\partial^{2} F}{\partial \tilde{z}_{1} \partial \tilde{z}_{2}} \approx 0$. Therefore equation \ref{eq:grad} can be re-written as 
\begin{align}
\gamma(\mathbf{x}) &= -\sum_{s}\sum_{r} \Bigg \langle
\begin{bmatrix}
E_{1} & 0 \\
0 & E{2}
\end{bmatrix}
\begin{bmatrix}
\frac{\partial^{2} F}{\partial \tilde{z}_{1} \partial v} \\ \frac{\partial^{2} F}{\partial \tilde{z}_{2} \partial v}
\end{bmatrix},  
\begin{bmatrix}
\Delta z_{1} \\ \Delta z_{2}
\end{bmatrix} \Bigg \rangle,  \nonumber \\
&=-\sum_{s}\sum_{r}\bigg(E_{1}\frac{\partial^{2} F}{\partial \tilde{z}_{1} \partial v}\Delta z_{1}+E_{2}\frac{\partial^{2} F}{\partial \tilde{z}_{2} \partial v}\Delta z_{2}\bigg), 
\label{eq:grad}
\end{align}
where $E_{1}=\frac{1}{\frac{\partial^{2} F}{\partial \tilde{z}_{1}^{2}}}$ and $E_{2}=\frac{1}{\frac{\partial^{2} F}{\partial \tilde{z}_{2}^{2}}}$ are the weighting parameters. For a more general $n$-dimensional case, its gradient formula can be found by extending equation \ref{eq:grad} to $n$-dimensions

\begin{align}
\gamma(\mathbf{x}) &= -\sum_{s}\sum_{r} \Bigg \langle
\begin{bmatrix}
\frac{\partial^{2} F}{\partial z_{1}^{2}} & \frac{\partial^{2} F}{\partial z_{1}\partial z_{2} }  &  \frac{\partial^{2} F}{\partial z_{1}\partial z_{3} } & \ldots & \frac{\partial^{2} F}{\partial z_{1}\partial z_{n} }  \\
 \frac{\partial^{2} F}{\partial z_{2}\partial z_{1} } & \frac{\partial^{2} F}{\partial z_{2}^{2}}  &  \frac{\partial^{2} F}{\partial z_{2}\partial z_{3} }  & \dots  & \frac{\partial^{2} F}{\partial z_{2}\partial z_{n} }\\
 \frac{\partial^{2} F}{\partial z_{3}\partial z_{1} }  & \frac{\partial^{2} F}{\partial z_{3}\partial z_{2} }  & \frac{\partial^{2} F}{\partial z_{3}^{2} } & \ldots & \frac{\partial^{2} F}{\partial z_{3}\partial z_{n} } \\ 
 \vdots & \vdots & \vdots & \ddots & \vdots \\
 \frac{\partial^{2} F}{\partial z_{n}\partial z_{1} } & \frac{\partial^{2} F}{\partial z_{n}\partial z_{2} } & \frac{\partial^{2} F}{\partial z_{n}\partial z_{3} } & \dots & \frac{\partial^{2} F}{\partial z_{n}^2 }
\end{bmatrix}^{-1}
\begin{bmatrix}
\frac{\partial^{2} F}{\partial z_{1}\partial v(x)} \\
\frac{\partial^{2} F}{\partial z_{2}\partial v(x)} \\
\frac{\partial^{2} F}{\partial z_{3}\partial v(x)} \\
\vdots \\
\frac{\partial^{2} F}{\partial z_{n}\partial v(x)}
\end{bmatrix},
\begin{bmatrix}
\Delta z_{1} \\ \Delta z_{2} \\ \Delta z_{3} \\ \vdots \\ \Delta z_{n}
\end{bmatrix} \Bigg \rangle, \nonumber \\
&=-\sum_{s}\sum_{r}\sum_{k} E_{k}\frac{\partial^{2} F}{\partial \tilde{z}_{k} \partial v}\Delta z_{k} \nonumber, \\
&=-\sum_{s}\sum_{r} \int \frac{\partial p_{(z_{1},z_{2})}^{syn}(\mathbf{x}_{r};\mathbf{x}_{s})}{\partial v} \Big( \sum_{k}\frac{\partial p_{(z_{1}-\Delta z_{1},z_{2}-\Delta z_{2})}^{obs}(\mathbf{x}_{r};\mathbf{x}_{s})}{\partial \tilde{z}_{k}}E_{k}\Delta z_{k} \Big)   dt.
\label{eq:grad_nd}
\end{align}
The Fr\'{e}chet derivative $\frac{\partial p}{\partial v}$ of the first-order acoustic equation can be written as \citep{chen2020seismic} 
\begin{equation}
\frac{\partial p}{\partial v}=-2\rho v g_{p}(\mathbf{x}_{r},t;\mathbf{x},0) * \nabla \cdot \mathbf{v}(\mathbf{x},t;\mathbf{x}_{s}),
\label{eq:frechet}
\end{equation} 
where $\rho$ and $v$ represent the density and velocity, respectively. The particle velocity is indicated by $\mathbf{v}$, $g_{p}$ is the Green's function, and $*$ means temporal convolution. Substituting equation \ref{eq:frechet} into equation \ref{eq:grad_nd} allows the MNML gradient to be expressed as
\begin{equation}
\gamma(\mathbf{x})=\sum_{s}\sum_{r}\int dt \big( 2\rho v g_{p}(\mathbf{x}_{r},t;\mathbf{x},0) * \nabla \cdot \mathbf{v}(\mathbf{x},t;\mathbf{x}_{s}) \big)\Delta p_{\mathbf{z}}(\mathbf{x},t;\mathbf{x}_{s}),
\label{eq:grad_final}
\end{equation} 
where $\Delta p_{\mathbf{z}}(\mathbf{x},t;\mathbf{x}_{s})=\sum_{k}\frac{\partial p_{(z_{1}-\Delta z_{1},z_{2}-\Delta z_{2})}^{obs}(\mathbf{x}_{r};\mathbf{x}_{s})}{\partial \tilde{z}_{k}}E_{k}\Delta z_{k}$ denotes the virtual source at the receiver location. For each LS dimension, a local virtual source is computed by weighting its LS feature difference $\Delta z_{k}$ on the partial derivative $\frac{\partial p_{(z_{1}-\Delta z_{1},z_{2}-\Delta z_{2})}^{obs}(\mathbf{x}_{r};\mathbf{x}_{s})}{\partial \tilde{z}_{k}}$, and then re-scale its value by dividing the weighting parameter $E_{k}$. The final virtual source $\Delta p_{\mathbf{z}}(\mathbf{x},t;\mathbf{x}_{s})$ can be obtained by summing all the local virtual sources at each LS dimensions together. Once we have the gradient, the velocity model can be updated by using the steepest descent formula 
\begin{equation}
v(\mathbf{x})_{k+1}=v(\mathbf{x})_{k}+\alpha_{k}\gamma(\mathbf{x})_{k},
\label{eq:update}
\end{equation}
where $k$ represents the iteration index and $\alpha_{k}$ indicates the step length. 

\subsection{Workflow of MNML}
\begin{figure}
\centering
\includegraphics[width=0.9\columnwidth]{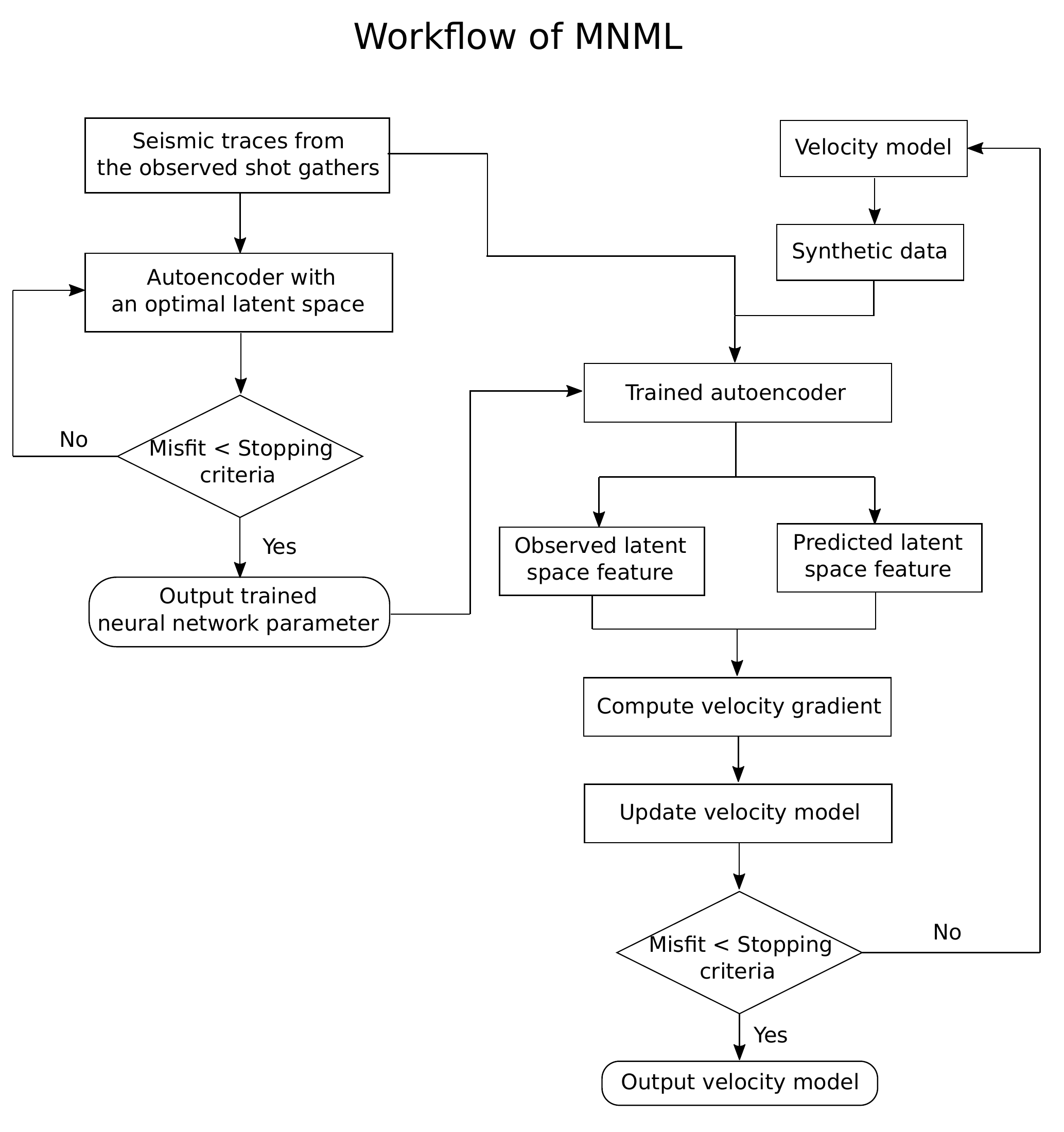}
\caption{The workflow of the multi-dimensional Newtonian machine learning inversion.}
\label{fig:workflow2}
\end{figure}

The workflow of the MNML inversion is shown in Figure \ref{fig:workflow2}, where an autoencoder is  trained to generate multi-dimensional LS features of the seismic data. The seismic traces in the observed shot gathers are often used to train and validate the autoencoder network. This autoencoder is only trained once and its parameters are fixed during the MNML inversion. At each iteration of the MNML inversion, we feed the observed and predicted data to the well-trained autoencoder to get their LS features. We then compute the velocity gradient using equation \ref{eq:grad_final} and update the current velocity model. We stop the inversion when the LS feature misfit goes below a certain threshold. 

\section{Numerical Tests}
We now use both synthetic and field data to test the effectiveness of the MNML method in recovering a high-resolution velocity model. We also compare the MNML results with the FWI results to demonstrate that the image resolution recovered by the MNML method is similar to FWI. 

\subsection{Checkerboard tests}
We first test the MNML method on the data generated by checkerboard models with three different acquisition geometries to demonstrate the capability of the MNML method in dealing with different acquisition systems.

\subsubsection{Crosswell checkerboard test}
\begin{figure}[!h]
\centering
\includegraphics[width=0.80\columnwidth]{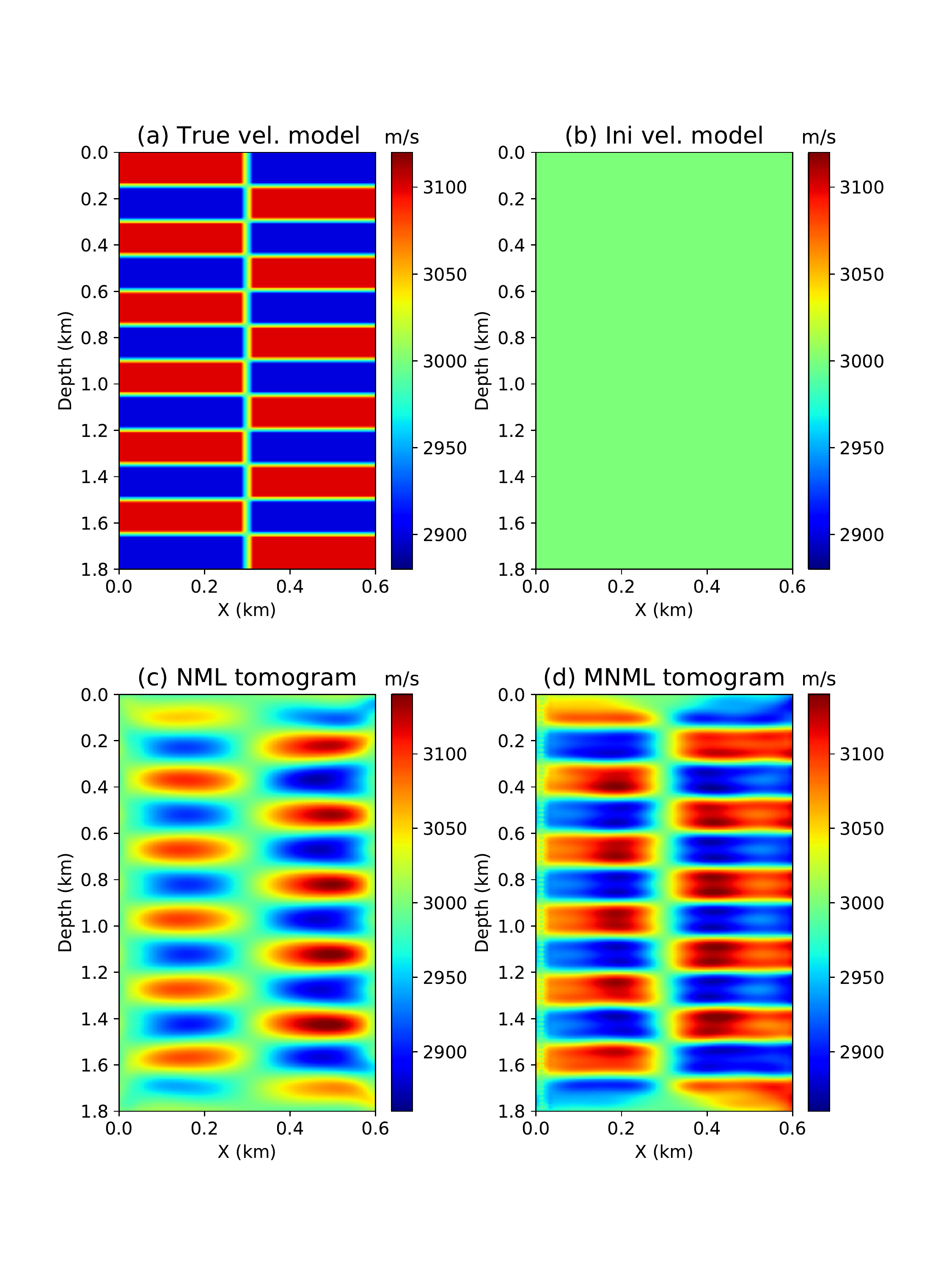}
\caption{(a) The true and (b) initial velocity models. (c) The NML and (d) MNML inverted velocity models.}
\label{fig:check1}
\end{figure}

Figures \ref{fig:check1}a and \ref{fig:check1}b show the true and initial velocity models for the checkerboard test with a crosswell geometry. The source well is at x = 10 m with 89 shots evenly deployed along the well. Each shot contains 179 receivers that are equally distributed on the receiver well at x = 590 m. The finite-difference modeling method is used to generate the seismic data and a 15 Hz Ricker wavelet is used as the seismic source. The NML method is first used to recover the velocity model. The NML inverted result is shown in Figure \ref{fig:check1}c where the low-wavenumber velocity information has been well recovered, but the high-wavenumber details are missing. The MNML method uses the NML tomogram as the initial model and inverts for the high-wavenumber velocity details by using higher-dimensional LS features. Figure \ref{fig:check1}d shows the MNML inverted model where the high-wavenumber velocity details are well recovered.  

\subsubsection{Early arrival surface geometry checkerboard test}
\begin{figure}[!h]
\centering
\includegraphics[width=1.0\columnwidth]{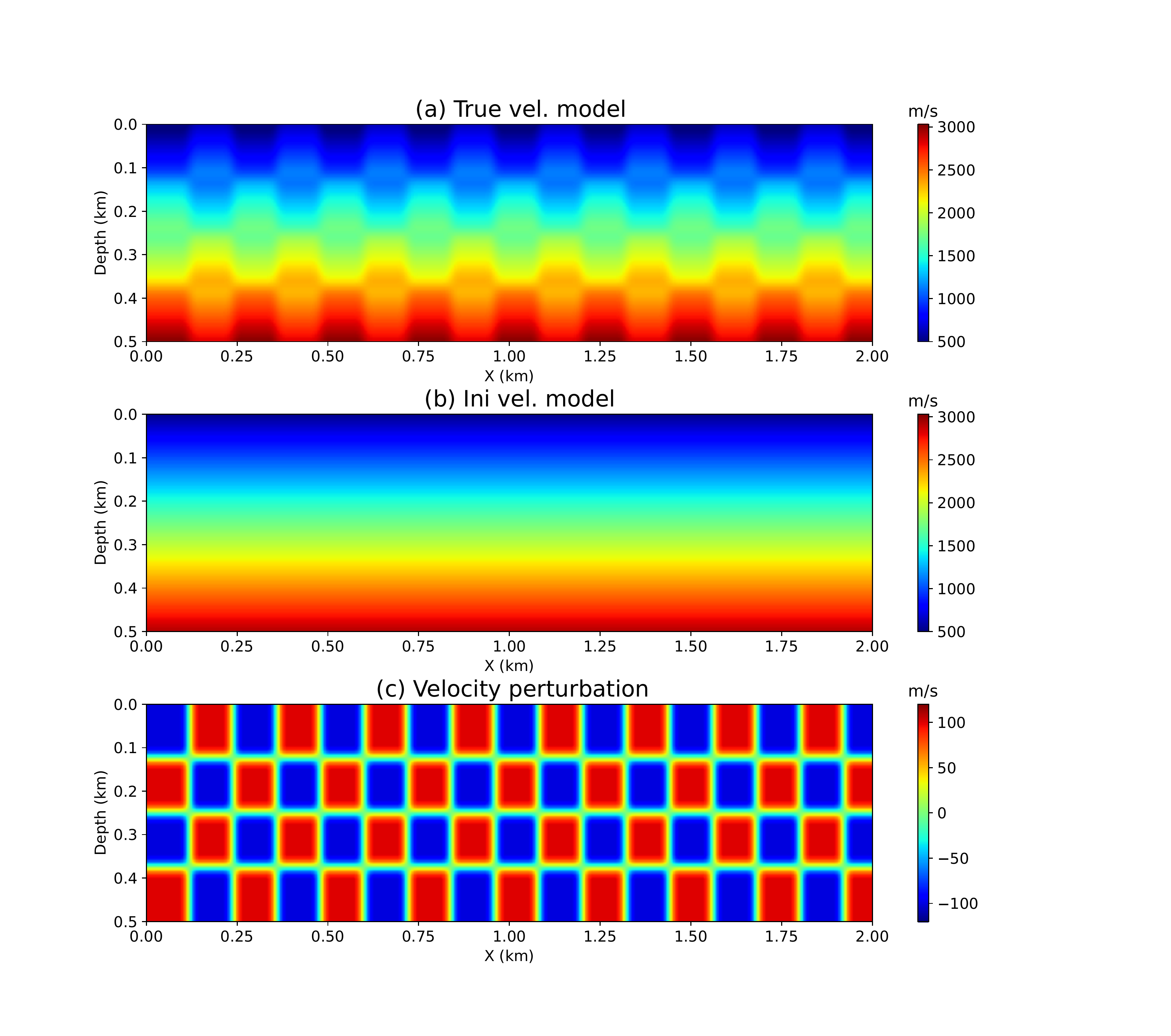}
\caption{(a) The true and (b) initial velocity models. (c) The perturbation model computed by doing subtraction between the true and initial model.}
\label{fig:check2}
\end{figure}

\begin{figure}[!h]
\centering
\includegraphics[width=1.0\columnwidth]{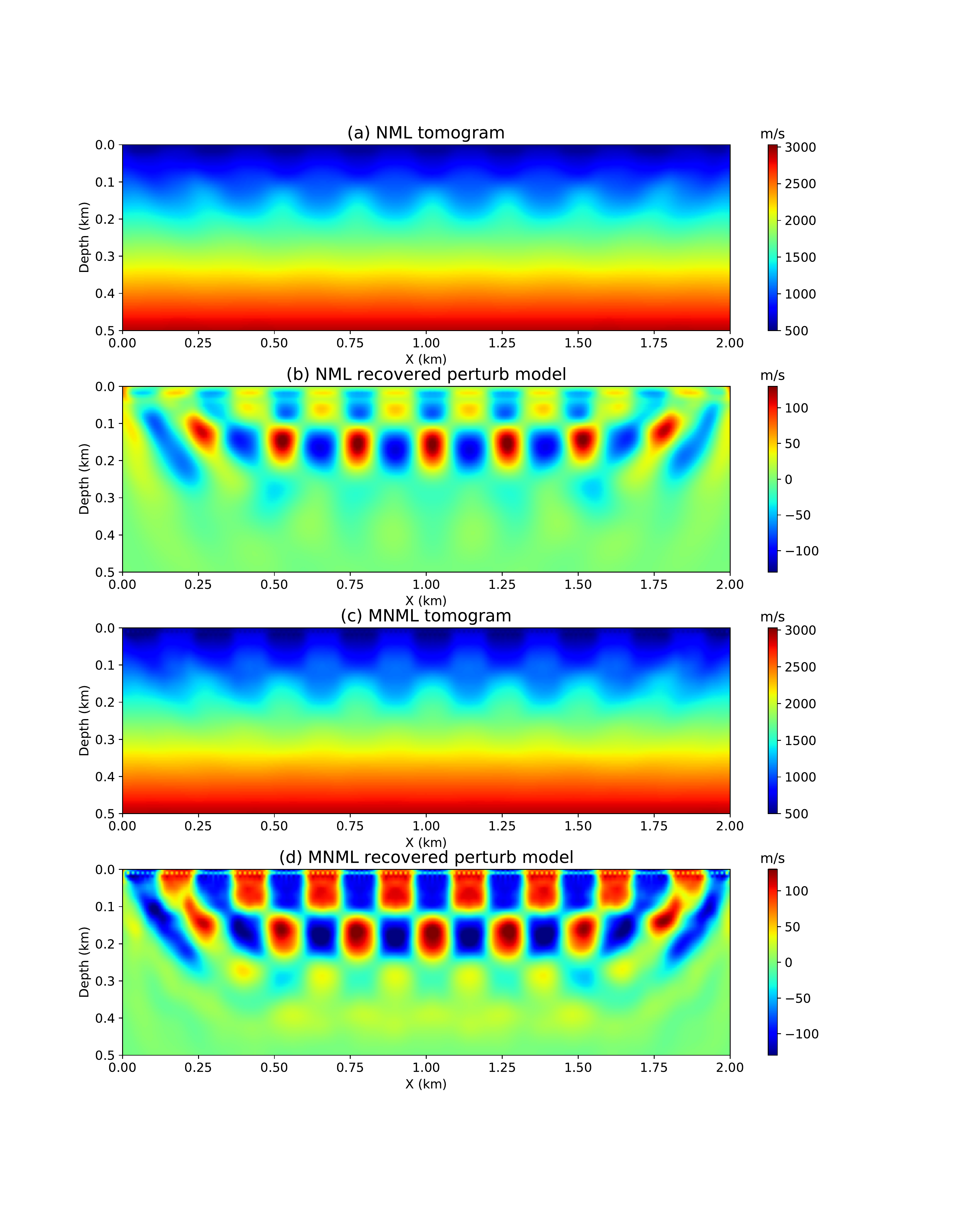}
\caption{(a) The NML recoverd velocity and (b) perturbation models. (c) The MNML recovered velocity and (d) perturbation model.}
\label{fig:check3}
\end{figure}

In the second checkerboard test, the sources and receivers are evenly deployed on the surface with fixed intervals of 16 m and 8 m, respectively. The true and initial velocity models are shown in Figure \ref{fig:check2}a and \ref{fig:check2}b. The velocity perturbation shows in Figure \ref{fig:check2}c is computed by doing the subtraction between the true and initial velocity models, which is the aim of the NML and MNML inversion. The source function used in the finite-difference modeling is a Ricker wavelet with a peak frequency of 12 Hz. Figures \ref{fig:check3}a and \ref{fig:check3}b show the velocity model and velocity perturbations recovered by the NML method. Similar to the previous test, the NML result mainly contains the low-wavenumber velocity information. However, the velocity model and velocity perturbations recovered by the MNML method given in Figure \ref{fig:check3}c and \ref{fig:check3}d have recovered more high-resolution velocity details.

\subsubsection{Reflection energy surface geometry checkerboard test}
\begin{figure}
\centering
\includegraphics[width=0.75\columnwidth]{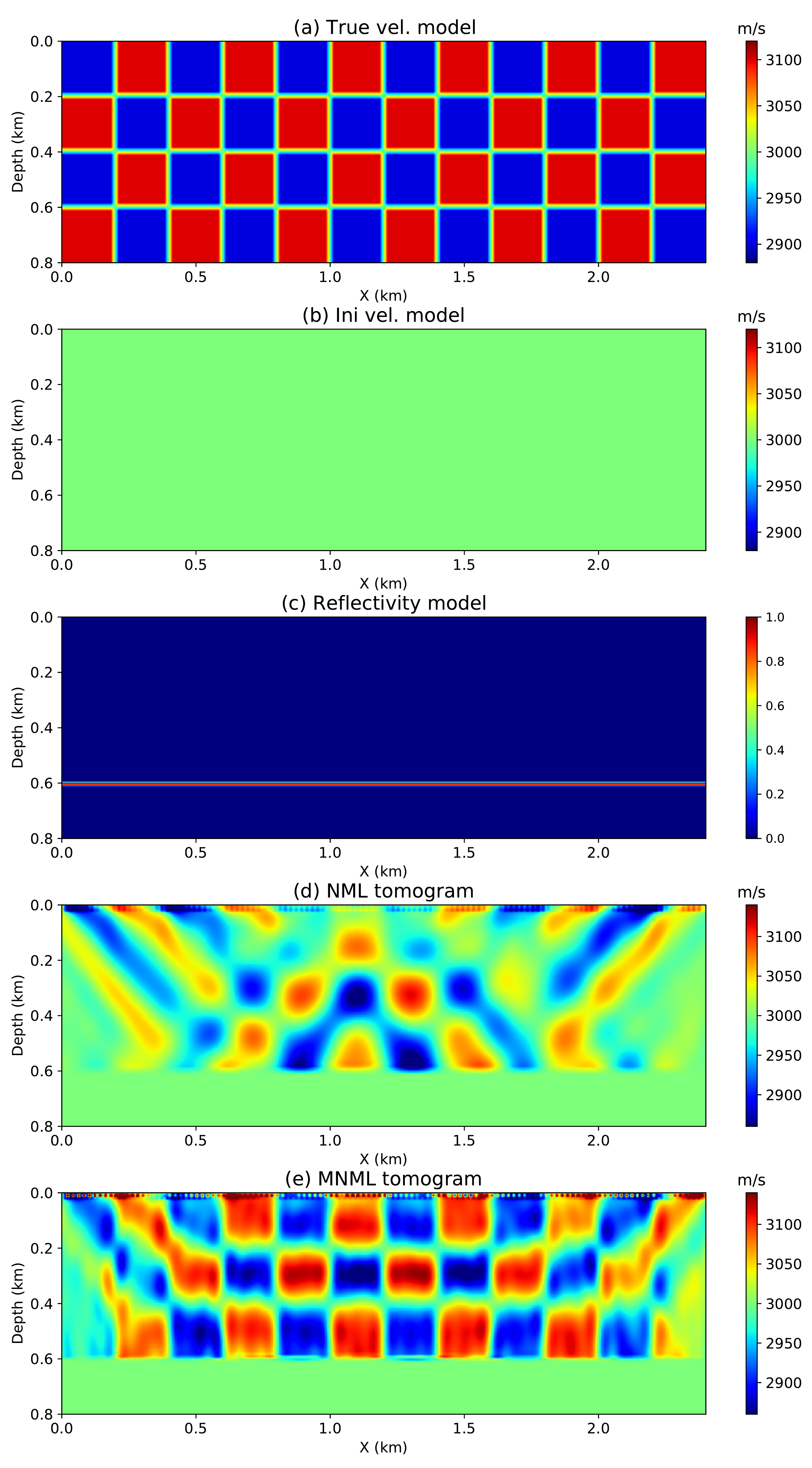}
\caption{(a) The true and (b) initial velocity models. (c) The reflectivity model used in Born modeling. (d) The NML and (e) MNML recovered velocity model.}
\label{fig:check4}
\end{figure}

The surface acquisition geometry is still employed in this test, but a reflection inversion engine is used for velocity inversion. The Born modeling method is used to generate the observed data using the true velocity and reflectivity models shown in Figures \ref{fig:check4}a and \ref{fig:check4}c, respectively. The observed dataset contains 119 shot gathers, each shot gather has 239 receivers. The sources and receivers are evenly distributed on the surface with source and receiver intervals of 20 m and 10 m, respectively. A 15 Hz Ricker wavelet is used as the source wavelet. The initial model given in Figure \ref{fig:check4}b is a homogeneous model with a velocity equals to 3000 m/s. The NML method is first used to recover the subsurface velocity model. As we expected, the NML tomogram shown in Figure \ref{fig:check4}d only recovered the low-wavenumber velocity information. Based on the NML tomogram, the MNML result given in Figure \ref{fig:check4}e has well recovered more high-wavenumber velocity details. Therefore, the whole velocity content can be well recovered by the NML and MNML methods working together. 

\subsection{Pluto model}
\begin{figure}[!h]
\centering
\includegraphics[width=1.0\columnwidth]{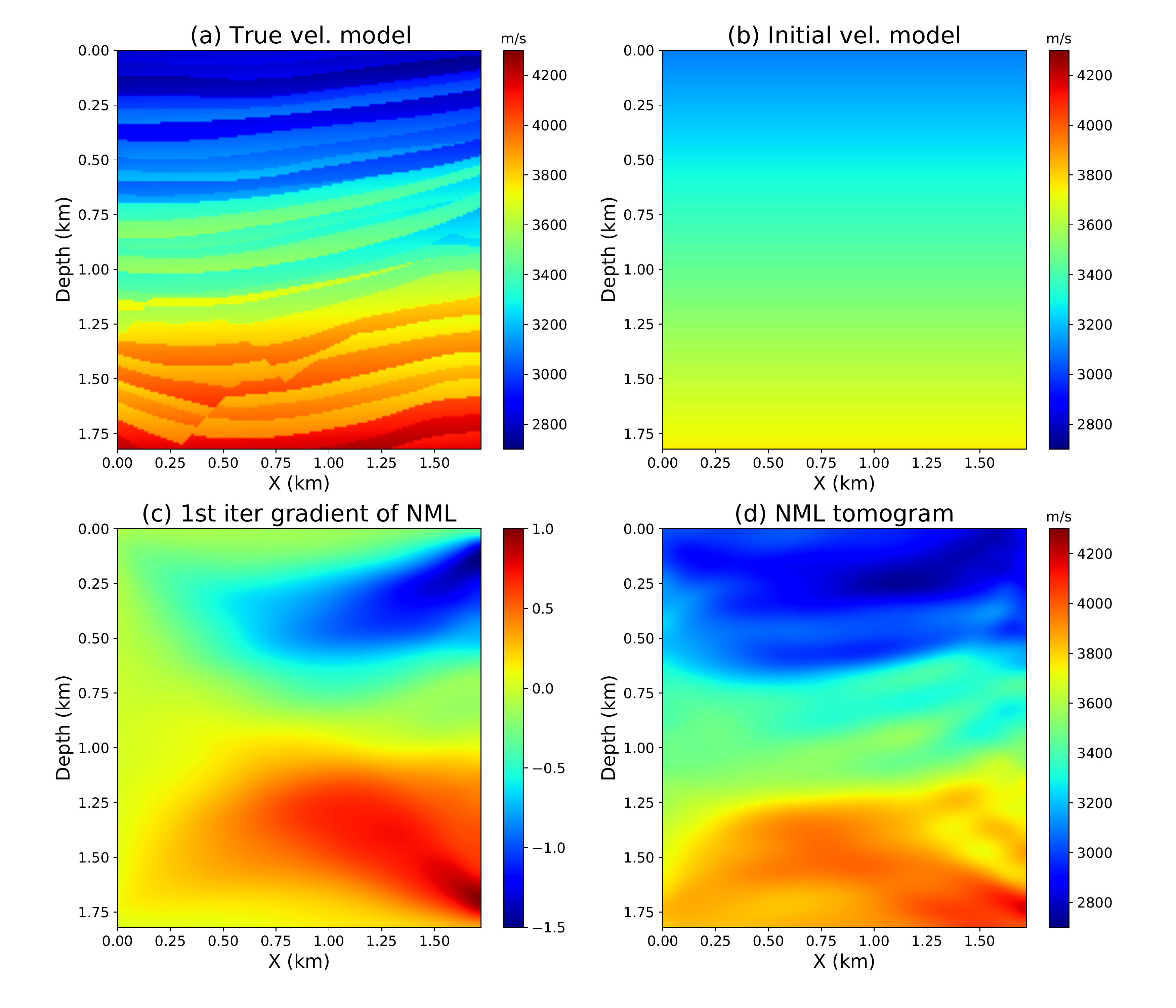}
\caption{(a) The true and (b) initial velocity models. (c) The first iteration gradient and (d) inverted velocity model of the NML method.}
\label{fig:pluto1}
\end{figure}

A portion of the Pluto model is selected as the true velocity model (shown in Figure \ref{fig:pluto1}a) to test the MNML method. A crosswell acquisition system is used to generate the seismic data. The source well is deployed at x=10 m which contains 59 shots at an equal interval of 30 m. Each shot has 177 receivers evenly distributed in the receiver well, which is located at x = 1750 m. A 20-Hz Ricker wavelet is used as the seismic source, and a linearly increasing velocity model is used as the initial velocity model. The minimum and maximum velocity of the initial model are equal to 3100 m/s and 3700 m/s, respectively. The acoustic finite-difference modeling is used to generate the observed data based on the true model. 

\subsubsection{NML inversion}
We first use the NML method to reconstruct the background velocity model. A three-layer autoencoder with a one-dimensional LS is designed to extract the one-dimensional LS feature from the seismic data. The first, second, and third encoder layers have dimensions of $1200\times1$, $200\times1$, and $15\times1$, respectively. The decoder network architecture is the mirror of the encoder network. The training dataset is composed of the envelope of the observed traces. During the training, the dataset is randomly separated into more than 200 mini-batches with a size of 50. The Adam optimizer is used to iteratively update the network. We stop the training when the reconstruction error barely decreases anymore, and achieved a training and validation accuracy of $97.51 \%$ and $97.17\%$, respectively. Figures \ref{fig:pluto1}c and \ref{fig:pluto1}d show the first iteration gradient and inverted velocity model of the NML method, which are dominated by the low-wavenumber velocity update. The NML inverted background model is used as the initial model for the MNML and FWI methods in the next step of inversion.  

\subsubsection{MNML inversion}
In this section, we use the MNML method to recover the high-wavenumber velocity details based on the NML inverted background model. We build a new autoencoder that has the same encoder and decoder network as the previous autoencoder. But we set its LS dimension to five (la=5), which is the optimal LS dimension found by looking at the ''data reconstruction error versus LS dimension" curve. The training dataset contains the original observed seismic traces, and the same training strategy is used for the autoencoder (la=5) training. The reason we use the original seismic data for training rather than its envelope is that the ten-dimensional LS feature has a stronger capability in representing complex features in the seismic data. 

\begin{figure}
\centering
\includegraphics[width=0.85\columnwidth]{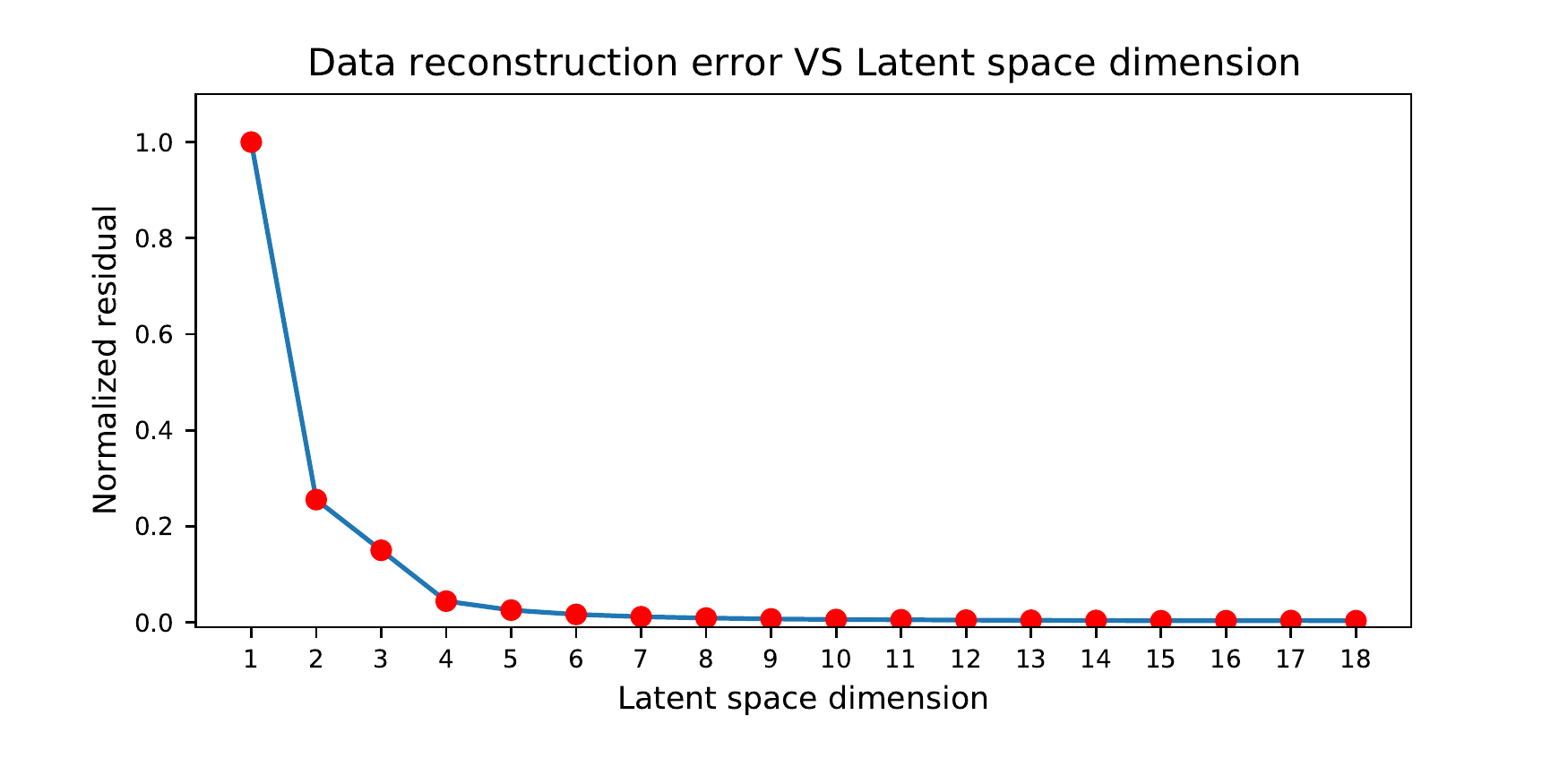}
\caption{The normalized data reconstruction error versus latent space dimension. The black arrow points to the optimal latent space dimension.}
\label{fig:recon_pluto}
\end{figure}	

\begin{figure}
\centering
\includegraphics[width=0.9\columnwidth]{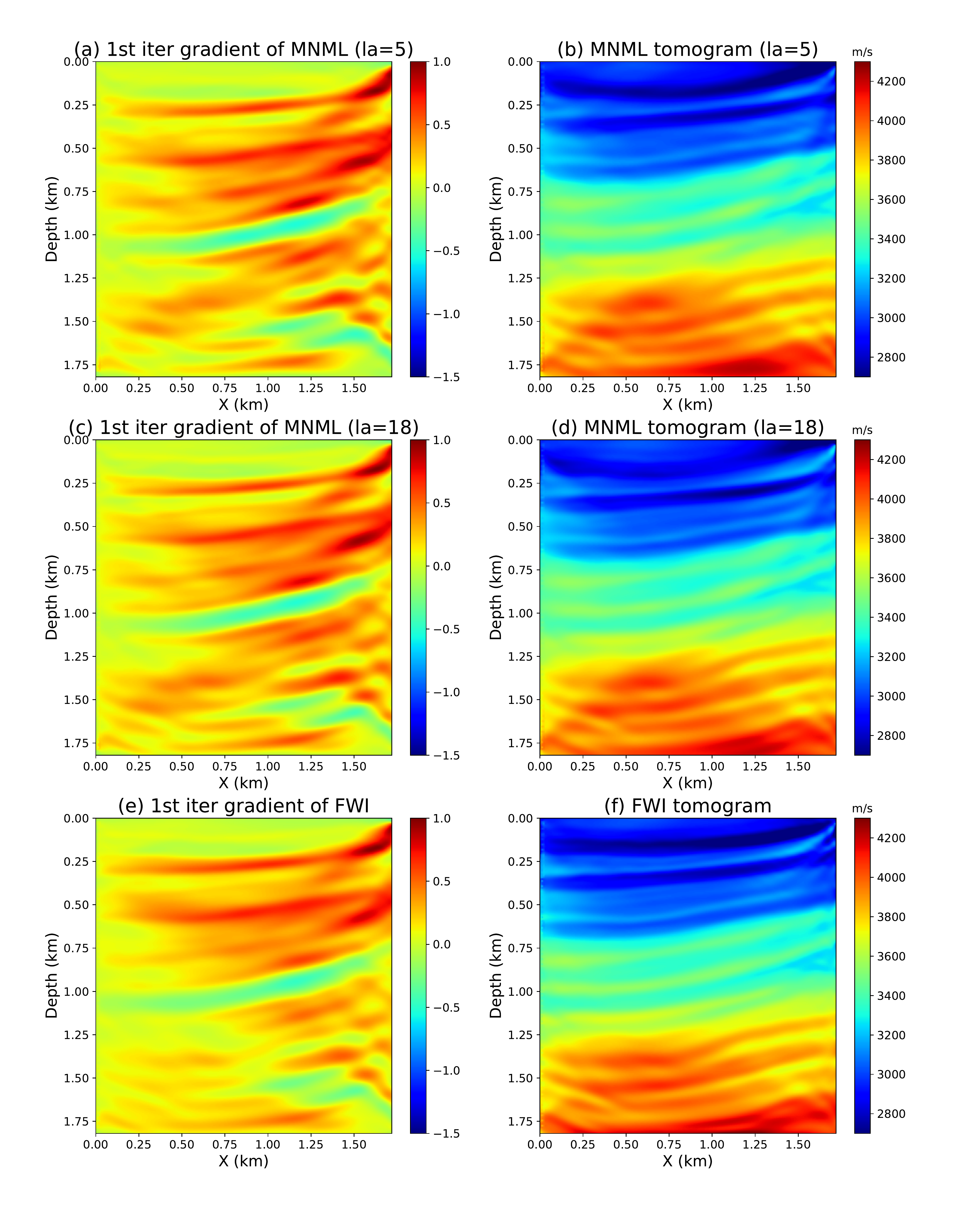}
\caption{(a) The first iteration gradient and (b) MNML tomogram (la=5), where the NML background tomogram is used as the starting model. (c) The first iteration gradient and (d) MNML tomogram (la=18). Here the MNML tomogram (la=5) is used as the starting model. (e) The first iteration gradient and (f) FWI tomogram, where the NML background tomogram is used as the starting model.}
\label{fig:pluto2}
\end{figure}

Figure \ref{fig:pluto2}a shows the first iteration gradient of MNML (la=5) which is dominated by the high-wavenumber updates. The inverted MNML tomogram (la=5) given in Figure \ref{fig:pluto2}b shows a dramatic resolution increase compared to the NML result. This suggests that the six-dimensional LS features contain more model information from the seismic data than the one-dimensional LS feature. To demonstrate that the six-dimensional LS is the optimal LS, we build another autoencoder with an eighteen-dimensional LS (la=18) and repeat the MNML inversion workflow. The MNML (la=18) gradient and tomogram given in Figures \ref{fig:pluto2}b and \ref{fig:pluto2}b are very similar to the MNML (la=5) results, which means the six-dimensional LS features are sufficient to recover the high-wavenumber velocity details of the model. 

To quantify the MNML inverted result, we use the FWI method to invert the seismic dataset and obtained a high-resolution model which is used as a benchmark model. Figures \ref{fig:pluto2}e and \ref{fig:pluto2}f show the FWI gradient after the 1st iteration and the FWI tomogram, which has the same level of image resolution compared to the MNML (la=5) tomogram. This suggests that the resolution of the MNML method is comparable to the FWI method. 

\subsection{Friendswood field data}
\begin{figure}
\centering
\includegraphics[width=0.75\columnwidth]{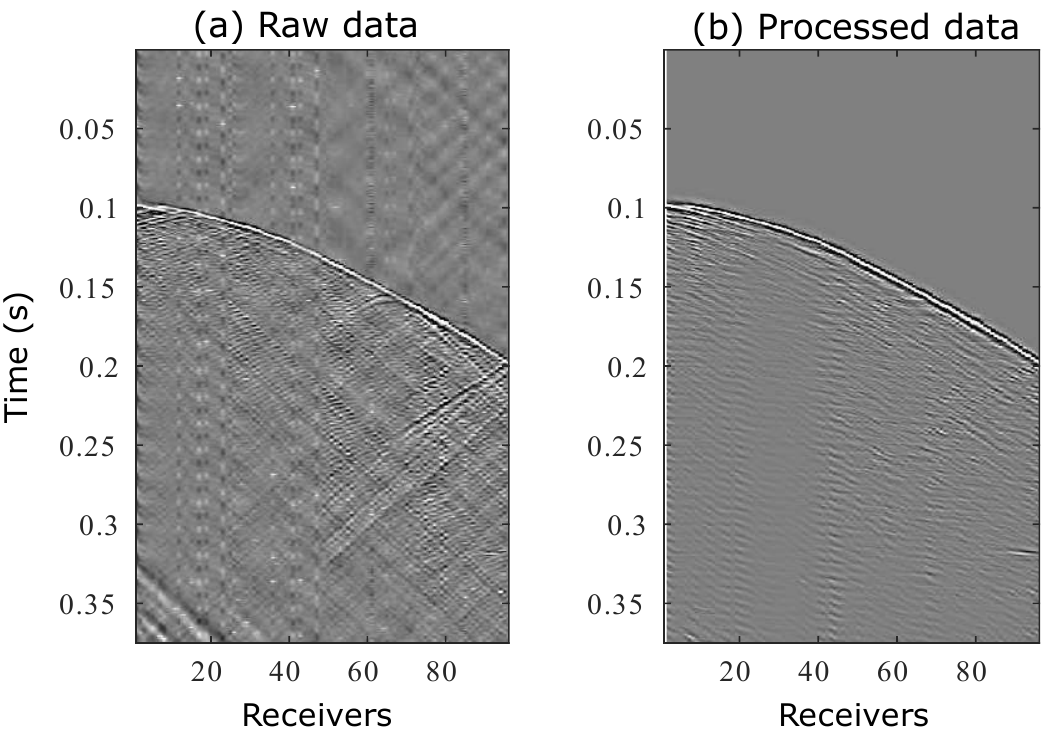}
\caption{(a) A raw shot gather recorded by a crosswell survey in Texas, US, and (b) its processed result.}
\label{fig:fri1}
\end{figure}

The field data test uses a crosswell dataset collected by Exxon in Texas \citep{chen1990subsurface}. The source and receiver wells are 305 m deep and separated by 183 m from each other. There are 97 shots fired at a shot interval of 3 m from z=9 m to z=305 m using downhole explosive charges. Each shot has 96 receivers distributed in the receiver well \citep{chen2020seismic}. The seismic data are recorded for 0.375 s with a time interval of 0.25 ms. The data contains extreme noise and tube waves.  We first bandpass the data to 80-400 Hz to remove the extreme linear noise. We then use a nine-point median filter to remove the tube wave. Finally, we separate the up and down-going waves using FK filtering. A similar processing workflow with more details can be found in \cite{cai1993processing, dutta2014attenuation, chen2020seismic}. We plot one example of the raw and processed data in Figures \ref{fig:fri1}a and \ref{fig:fri1}b, respectively.

\begin{figure}
\centering
\includegraphics[width=0.9\columnwidth]{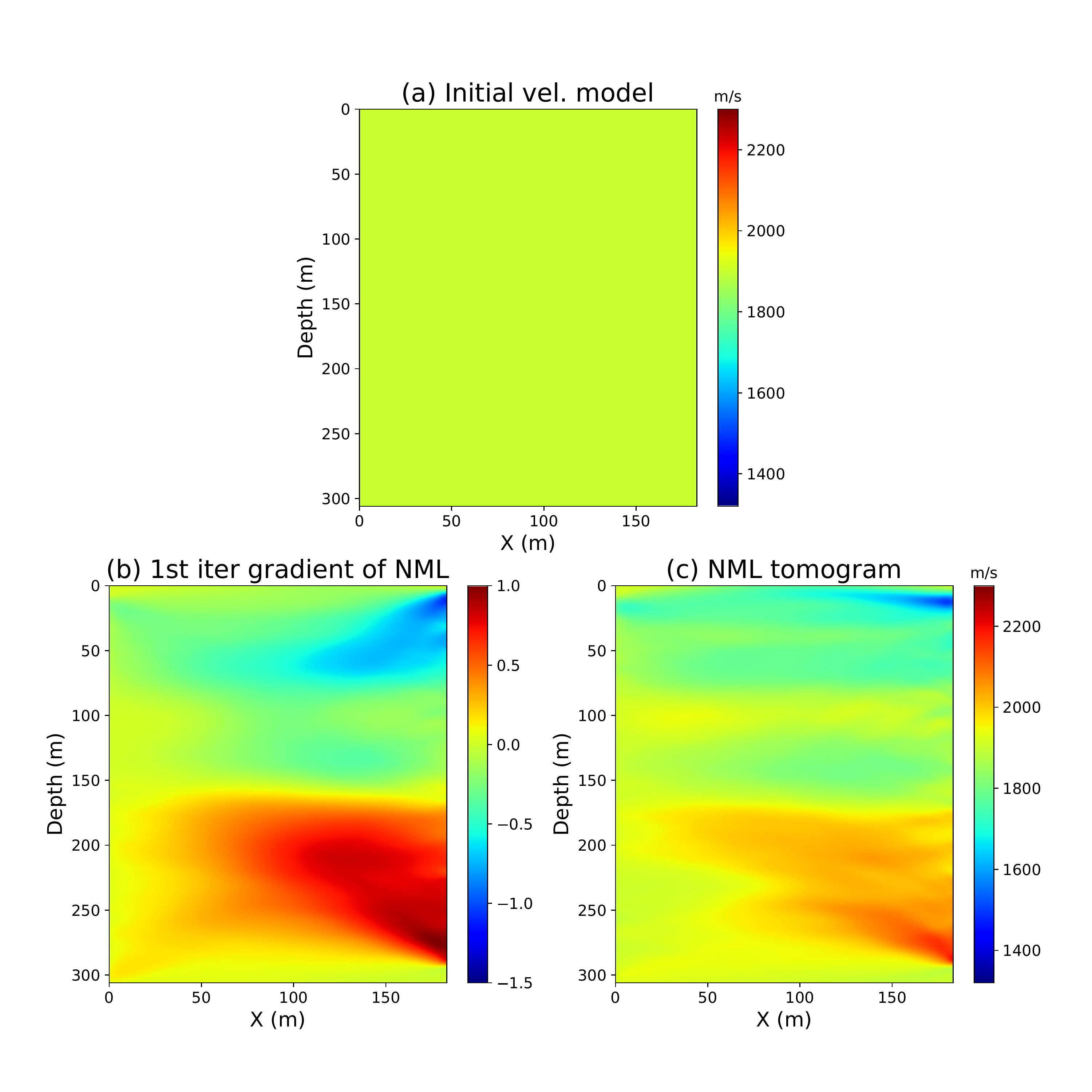}
\caption{(a) The initial homogeneous velocity model (vel = 1900 m/s). (b) The first iteration gradient and (c) NML tomogram.}
\label{fig:fri2}
\end{figure}

We assume the initial model is a homogeneous model with a velocity equals to 1900 m/s. We first use the NML method to recover the background velocity model. We design an autoencoder with four encoder and decoder layers and a single-dimensional LS. The first to the fourth encoder layers have a dimension of $3750 \times 1$, $1000\times1$, $200\times1$, $20\times1$, and the decoder network is the mirror of the encoder network. The training dataset contains the envelope of the observed seismic traces. We use the same training strategy described above to train this autoencoder. Figures \ref{fig:fri2}c and \ref{fig:fri2}d show the first iteration gradient and NML tomogram, which mainly contains the low-wavenumber velocity information. 

\begin{figure}
\centering
\includegraphics[width=0.85\columnwidth]{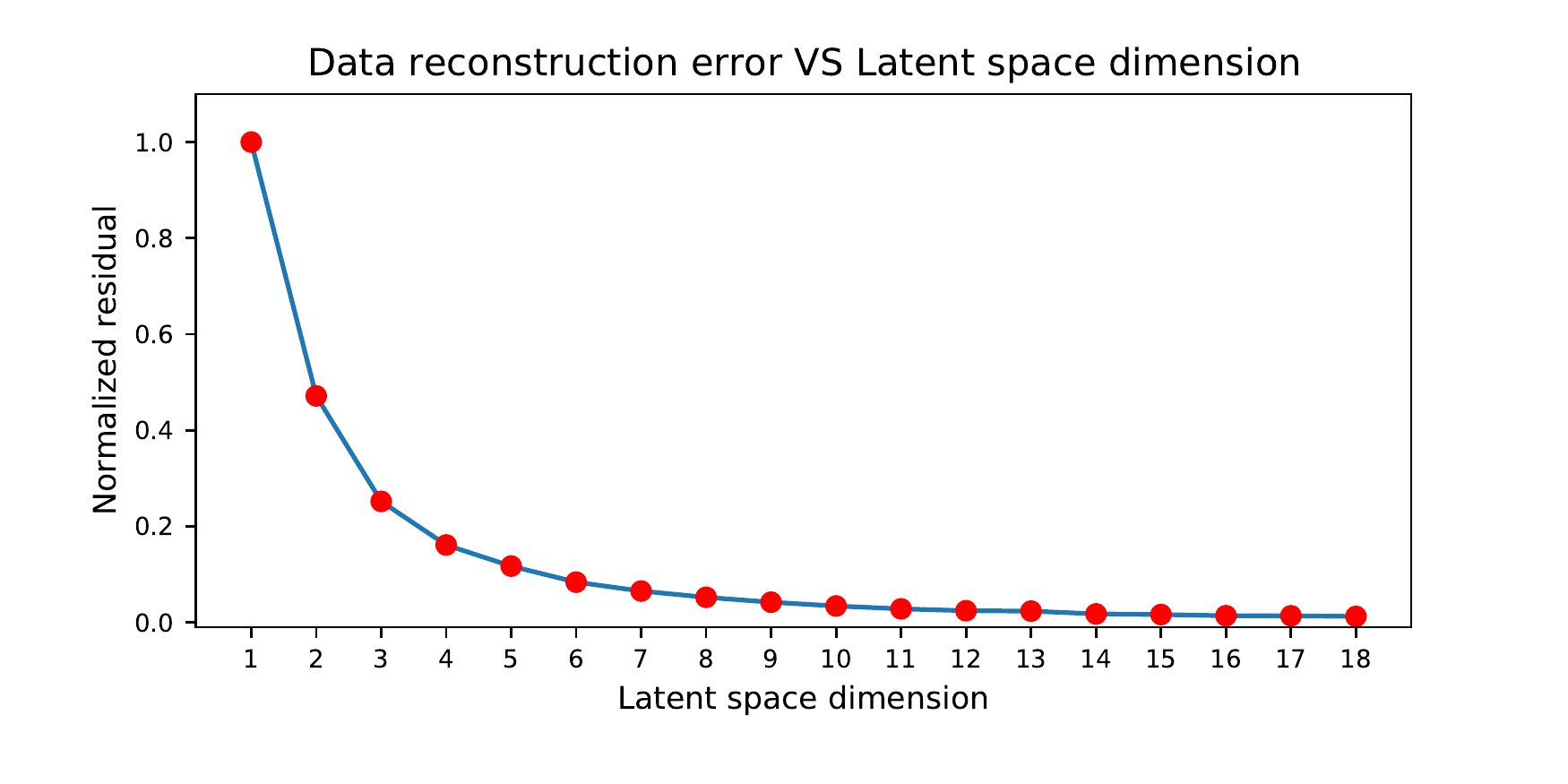}
\caption{The normalized data reconstruction error versus latent space dimension. The black arrow points to the optimal latent space dimension.}
\label{fig:recon_friends}
\end{figure}

\begin{figure}[!h]
\centering
\includegraphics[width=0.9\columnwidth]{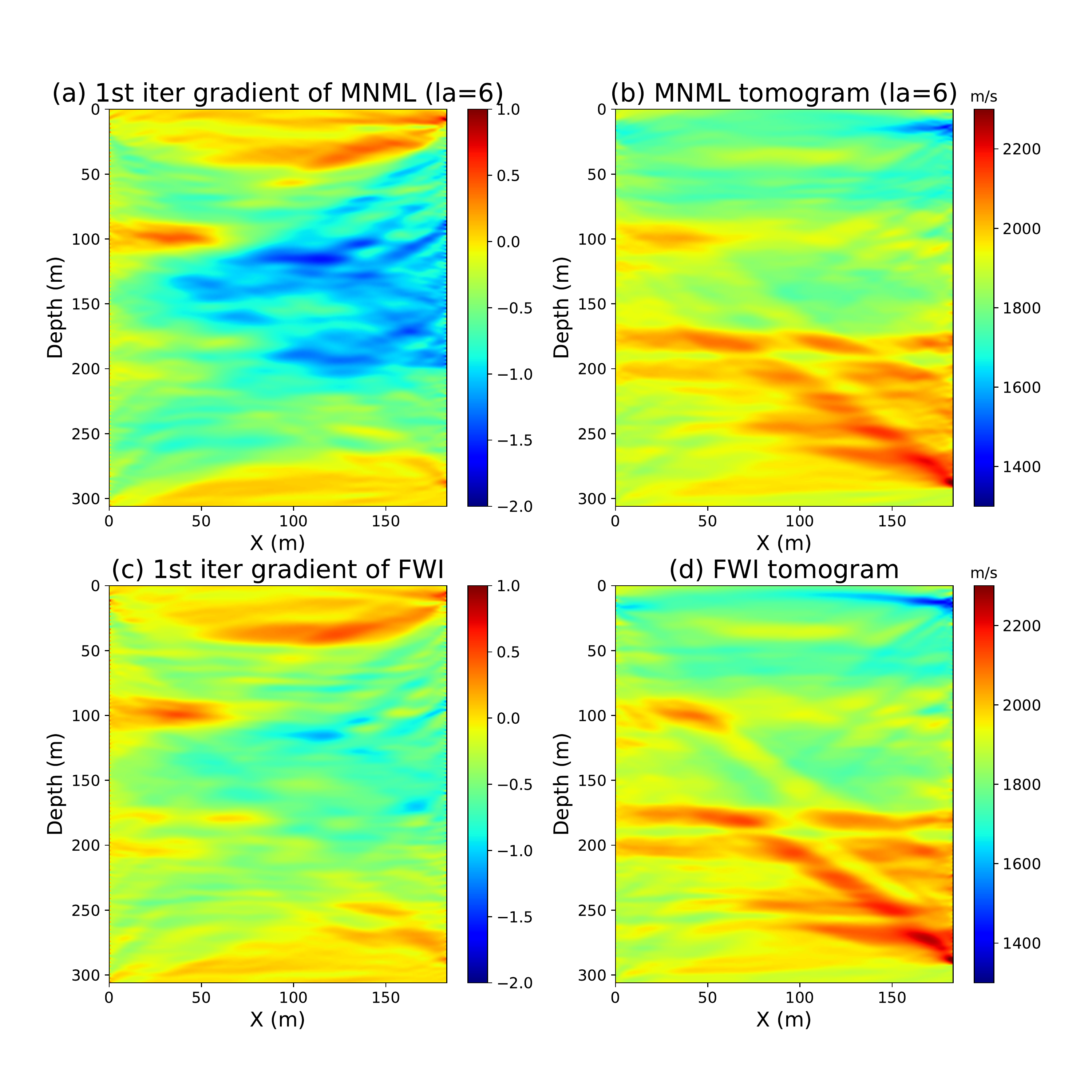}
\caption{(a) The first iteration gradient and (b) MNML tomogram (la=6). (c) The first iteration gradient and (d) FWI tomogram. Both these methods use the NML tomogram as the starting model. }
\label{fig:fri3}
\end{figure}

To reconstruct the high-wavenumber velocity details, we apply the MNML method on the NML background tomogram. We build a new autoencoder with a six-dimensional LS (la=6) but keep the encoder and decoder network the same as the previous autoencoder used in the NML. The training dataset for MNML includes the observed seismic traces rather than its envelopes. Figures \ref{fig:fri3}a and \ref{fig:fri3}b show the first iteration gradient and MNML tomogram (la=6), which contains the high-wavenumber velocity details. The MNML tomogram (la=6) shows a noticeable resolution increase compared to the NML tomogram. To validate the robustness of the MNML tomogram, we compare it with the FWI tomogram shown in Figure \ref{fig:fri3}d. Same as the MNML inversion, FWI also uses the NML background tomogram as the starting model. The FWI tomogram is almost identical to the MNML tomogram (la=6), which means the MNML tomogram is correct.

\section{Conclusion}
We present a multi-dimensional Newtonian machine learning (MNML) inversion method that inverts a multi-dimensional latent space (LS) feature to reconstruct the high-wavenumber velocity details of the subsurface model. An autoencoder with a multi-dimensional LS is trained to extract the low-dimensional representations from the high-dimensional seismic traces. The compressed data contains more information, such as the waveform variations, of the input seismic data than the one-dimensional LS. To compute the MNML gradient, a multi-variable connective function and the multi-variable implicit function theorem are used to mathematically connect the LS feature perturbations to the velocity perturbations. Numerical results show that the MNML method can recover the higher-wavenumber velocity information, and the resolution of the MNML tomogram is comparable to the FWI tomogram. However, MNML requires a much smaller storage space than FWI because only the low-dimensional representations of the seismic data need to be saved in the computer. 

A potential problem of MNML is that it may get stuck in the local minima compared to the NML method and conventional skeletonized inversions. This is mainly due to the reason that more nonlinear characteristics of the seismic data, such as the waveform variations, are contained in the multi-dimensional LS feature. So the MNML misfit function has more local minima than the NML misfit function. Therefore we suggest a multiscale inversion approach that gradually uses a higher dimensional LS feature for inversion.


\append{Multi-variable implicit function theorm}
The terms $\frac{\partial z_{1}}{\partial v(x)}$ and $\frac{\partial z_{2}}{\partial v(x)}$ in gradient $\gamma$ cannot be directly calculated  as there is no governing equation which includes both $z_{1}$, $z_{2}$ and velocity $v(x)$. However, the connective function in equation \ref{eq:connect1} implicitly connect $z_{1}$, $z_{2}$ with $v(x)$. Equation \ref{eq:connect2} can be rewritten as 
\begin{align}
\frac{\partial F}{\partial z_{1}}&=\bar{F_{1}}(z_{1},z_{2},v)=0,  \nonumber \\
\frac{\partial F}{\partial z_{2}}&=\bar{F_{2}}(z_{1},z_{2},v)=0, \label{eq:ap1}
\end{align}
where $\bar{f_{1}}$ and $\bar{f_{2}}$ are function of $z_{1}$, $z_{2}$ and $v$. Moreover, the latent space variables $z_{1}$ and $z_{2}$ are also the function of velocity $v$:
\begin{align}
z_{1} &= z_{1} (v(\mathbf{x})), \nonumber \\
z_{2} &= z_{2}(v(\mathbf{x})).  \label{eq:ap2}
\end{align}
Differentiating equation \ref{eq:ap1} we get
\begin{align}
\frac{\partial \bar{F_{1}}}{\partial z_{1}}d z_{1}+\frac{\partial \bar{F_{1}}}{\partial z_{2}}dz_{2}+\frac{\partial \bar{F_{1}}}{\partial v}dv&=0, \nonumber \\
\frac{\partial \bar{F_{2}}}{\partial z_{1}}d z_{1}+\frac{\partial \bar{F_{2}}}{\partial z_{2}}dz_{2}+\frac{\partial \bar{F_{2}}}{\partial v}dv&=0. \label{eq:ap3}
\end{align}
Then we differentiate equation \ref{eq:ap3} with respect to $v$ to obtain
\begin{align}
\frac{\partial \bar{F_{1}}}{\partial z_{1}}\frac{\partial z_{1}}{\partial v}+\frac{\partial \bar{F_{1}}}{\partial z_{2}}\frac{\partial z_{2}}{\partial v}+\frac{\partial \bar{F_{1}}}{\partial v}&=0, \nonumber \\
\frac{\partial \bar{F_{2}}}{\partial z_{1}}\frac{\partial z_{1}}{\partial v}+\frac{\partial \bar{F_{2}}}{\partial z_{2}}\frac{\partial z_{2}}{\partial v}+\frac{\partial \bar{F_{2}}}{\partial v}&=0. \label{eq:ap4}
\end{align}
Equation \ref{eq:ap4} can be re-written in matrix-vector multiplication form as 
\begin{align}
\begin{bmatrix}
\frac{\partial \bar{F_{1}}}{\partial z_{1}} & \frac{\partial \bar{F_{1}}}{\partial z_{2}}\\
\frac{\partial \bar{F_{2}}}{\partial z_{1}} & \frac{\partial \bar{F_{2}}}{\partial z_{2}}
\end{bmatrix}^{-1}
\begin{bmatrix}
\frac{\partial z_{1}}{\partial v(x)} \\ \frac{\partial z_{2}}{\partial v(x)}
\end{bmatrix}=
\begin{bmatrix}
\frac{\partial \bar{F_{1}}}{\partial v} \\ \frac{\partial \bar{F_{2}}}{\partial v}
\end{bmatrix} \label{eq:ap5}
\end{align}
To solve $\frac{\partial z_{1}}{\partial v}$ and $\frac{\partial z_{2}}{\partial v}$, we get 
\begin{align}
\begin{bmatrix}
\frac{\partial z_{1}}{\partial v(x)} \\ \frac{\partial z_{2}}{\partial v(x)}
\end{bmatrix}&=
\begin{bmatrix}
\frac{\partial \bar{F_{1}}}{\partial z_{1}} & \frac{\partial \bar{F_{1}}}{\partial z_{2}} \\
\frac{\partial \bar{F_{2}}}{\partial z_{1}} & \frac{\partial \bar{F_{2}}}{\partial z_{2}} 
\end{bmatrix}^{-1}
\begin{bmatrix}
\frac{\partial \bar{F_{1}}}{\partial v} \\
\frac{\partial \bar{F_{2}}}{\partial v}
\end{bmatrix} \nonumber \\
&=
\begin{bmatrix}
\frac{\partial^{2} F}{\partial z_{1}^{2}} & \frac{\partial^{2} F}{\partial z_{1} \partial z_{2}}  \\
\frac{\partial^{2} F}{\partial z_{2} \partial z_{1}} & \frac{\partial^{2} F}{\partial z_{2}^{2}} 
\end{bmatrix}^{-1}
\begin{bmatrix}
\frac{\partial^{2} F}{\partial z_{1}\partial v} \\ \frac{\partial^{2} F}{\partial z_{2}\partial v}
\end{bmatrix}
\end{align}

\append{Connective function derivation}
The optimal difference can be found by solving equation $\nabla F(\tilde{z}_{1},\tilde{z}_{2}) |_{(\tilde{z}_{1}=\Delta z_{1}, \tilde{z}_{2}=\Delta z_{2})}=0$ as
\begin{align}
\nabla F(\tilde{z}_{1},\tilde{z}_{2}) |_{(\tilde{z}_{1}=\Delta z_{1}, \tilde{z}_{2}=\Delta z_{2})} &=(\frac{\partial F}{\partial \tilde{z}_{1}},\frac{\partial F}{\partial \tilde{z}_{2}}) \nonumber \\
&=\int p_{(z_{1},z_{2})}^{syn}(\mathbf{x}_{r},t;\mathbf{x}_{s})\bigg (\frac{p_{(z_{1}-\Delta z_{1},z_{2}-\Delta z_{2})}^{obs}}{\partial \tilde{z}_{1}}, \frac{p_{(z_{1}-\Delta z_{1},z_{2}-\Delta z_{2})}^{obs}}{\partial \tilde{z}_{2}} \bigg) dt = 0 \label{eq:ap_connect2}
\end{align}
We can re-write equation \ref{eq:ap_connect2} as a system of equations
\begin{align}
\frac{\partial F}{\partial \tilde{z}_{1}} &= \int p_{(z_{1},z_{2})}^{syn} \frac{p_{(z_{1}-\Delta z_{1},z_{2}-\Delta z_{2})}^{obs}}{\partial \tilde{z}_{1}} dt=0   \nonumber \\
\frac{\partial F}{\partial \tilde{z}_{2}} &= \int p_{(z_{1},z_{2})}^{syn} \frac{p_{(z_{1}-\Delta z_{1},z_{2}-\Delta z_{2})}^{obs}}{\partial \tilde{z}_{2}} dt=0. \label{eq:ap_connect2_1}
\end{align}
Combing equation \ref{eq:ap_connect2_1} with the multi-variable implicit function theorem \citep{Guo2016multi}, we get the gradient formula for each LS dimension as
\begin{equation}
\begin{bmatrix}
\frac{\partial \Delta z_{1}}{\partial v} \\ \frac{\partial \Delta z_{2}}{\partial v}
\end{bmatrix}
=
\begin{bmatrix}
\frac{\partial^{2} F}{\partial \tilde{z}_{1}^{2}} & \frac{\partial^{2} F}{\partial \tilde{z}_{1} \partial \tilde{z}_{2}} \\
\frac{\partial^{2} F}{\partial \tilde{z}_{1} \partial \tilde{z}_{2}} & \frac{\partial^{2} F}{\partial \tilde{z}_{2}^{2}}
\end{bmatrix}^{-1}
\begin{bmatrix}
\frac{\partial^{2} F}{\partial \tilde{z}_{1} \partial v} \\ \frac{\partial^{2} F}{\partial \tilde{z}_{2} \partial v}
\end{bmatrix},
\label{eq:ap_connect3}
\end{equation}
where
\begin{align}
\frac{\partial^{2} F}{\partial \tilde{z}_{1} \partial v}&=\int \frac{\partial p_{(z_{1}-\Delta z_{1},z_{2}-\Delta z_{2})}^{obs}(\mathbf{x}_{r},t;\mathbf{x}_{s})}{\partial \tilde{z}_{1}}\frac{\partial p_{(z_{1},z_{2})}^{syn}(\mathbf{x}_{r},t;\mathbf{x}_{s})}{\partial v} dt \nonumber \\
\frac{\partial^{2} F}{\partial \tilde{z}_{2} \partial v}&=\int \frac{\partial p_{(z_{1}-\Delta z_{1},z_{2}-\Delta z_{2})}^{obs}(\mathbf{x}_{r},t;\mathbf{x}_{s})}{\partial \tilde{z}_{2}}\frac{\partial p_{(z_{1},z_{2})}^{syn}(\mathbf{x}_{r},t;\mathbf{x}_{s})}{\partial v} dt, \label{eq:ap_connect4}
\end{align}
and
\begin{align}
\frac{\partial^{2} F}{\partial \tilde{z}_{1}^{2}} &= \int \frac{\partial^{2} p_{(z_{1}-\Delta z_{1}, z_{2}-\Delta z_{2})}^{obs}(\mathbf{x}_{r},t;\mathbf{x}_{s})}{\partial \tilde{z}_{1}^{2}}p_{(z_{1},z_{2})}^{syn}(\mathbf{x}_{r},t;\mathbf{x}_{s})dt, \nonumber \\
\frac{\partial^{2} F}{\partial \tilde{z}_{2}^{2}} &= \int \frac{\partial^{2} p_{(z_{1}-\Delta z_{1}, z_{2}-\Delta z_{2})}^{obs}(\mathbf{x}_{r},t;\mathbf{x}_{s})}{\partial \tilde{z}_{2}^{2}}p_{(z_{1},z_{2})}^{syn}(\mathbf{x}_{r},t;\mathbf{x}_{s})dt, \nonumber \\
\frac{\partial^{2}F}{\partial \tilde{z}_{1} \partial \tilde{z}_{2}} &= \int \frac{\partial^{2} p_{(z_{1}-\Delta z_{1}, z_{2}-\Delta z_{2})}^{obs}(\mathbf{x}_{r},t;\mathbf{x}_{s})}{\partial \tilde{z}_{1} \partial \tilde{z}_{2} }p_{(z_{1},z_{2})}^{syn}(\mathbf{x}_{r},t;\mathbf{x}_{s})dt. \label{eq:ap_connect5}
\end{align}

\bibliographystyle{seg}  
\bibliography{paper}

\listoffigures

\end{document}